%% file: main.tex
\documentclass[sigconf, authorversion]{acmart}
\setcopyright{acmlicensed}
\copyrightyear{2026}
\acmYear{2026}
\acmDOI{XXXXXXX.XXXXXXX}
\acmISBN{978-1-4503-XXXX-X/2018/06}
\input{ethos}

\usepackage[T1]{fontenc}

\usepackage[epsilon,tsrm]{backnaur}
\usepackage{fontawesome}

\begin{document}

\title{\sys: Enabling Cross-Vendor Authentication for IoT}

\author{Sanket Goutam}
\affiliation{
  \institution{Stony Brook University}
  \city{Stony Brook, NY}
  \country{USA}
}
\email{sgoutam@cs.stonybrook.edu}

\author{Omar Chowdhury}
\affiliation{
  \institution{Stony Brook University}
  \city{Stony Brook, NY}
  \country{USA}
}
\email{omar@cs.stonybrook.edu}

\author{Amir Rahmati}
\affiliation{
  \institution{Stony Brook University}
  \city{Stony Brook, NY}
  \country{USA}
}
\email{amir@cs.stonybrook.edu}


\input{chapters/abstract}

\maketitle

\input{chapters/introduction}
\input{chapters/background}

\input{chapters/design}

\input{chapters/implementation}

\input{chapters/evaluation}
\input{chapters/discussion}

\input{chapters/related_works}
\input{chapters/conclusion}

\bibliographystyle{ACM-Reference-Format}
\bibliography{references}

\appendix

\section{Generative AI Usage Considerations}
We used GenAI tools, such as ChatGPT, to refine the text in this paper. The authors wrote the entire original draft and lightly used GenAI tools to polish the text, primarily focusing on the Introduction section. Any modified text was carefully reviewed for accuracy and correctness.

\input{chapters/appendix}

\end{document}

%% file: ethos.tex
\usepackage{xspace}
\usepackage{url}
\usepackage{multirow}
\def\baselinestretch{0.98} 

\usepackage{caption}
\usepackage{subfig}
\usepackage[symbol]{footmisc}

\let\OLDthebibliography\thebibliography
\renewcommand\thebibliography[1]{
  \OLDthebibliography{#1}
  \setlength{\parskip}{0pt}
  \setlength{\itemsep}{0pt plus 0.3ex}
}

\usepackage{pifont} 
\usepackage{amsmath}
\usepackage{todonotes} 
\usepackage{tikz}

\usepackage{enumitem} 
\setlist{nosep} 
\setlist{noitemsep,topsep=2pt,parsep=0pt,partopsep=0pt} 
\setlist[itemize]{leftmargin=*} 

\usepackage{wrapfig}
\usepackage{rotating} 

\usepackage{booktabs} 

\usepackage{array} 
\newcommand{\PreserveBackslash}[1]{\let\temp=\\#1\let\\=\temp}
\newcolumntype{C}[1]{>{\PreserveBackslash\centering}p{#1}}
\newcolumntype{R}[1]{>{\PreserveBackslash\raggedleft}p{#1}}
\newcolumntype{L}[1]{>{\PreserveBackslash\raggedright}p{#1}}

\makeatletter
\renewcommand\section{\@startsection{section}{1}{\z@}%
  {-5pt \@plus -2pt \@minus -2pt}
  {1pt \@plus 1pt \@minus 1pt}
  {\normalfont\large\bfseries}}
\renewcommand\subsection{\@startsection{subsection}{2}{\z@}%
  {-4pt \@plus -2pt \@minus -1pt}
  {0.5pt \@plus 0.5pt \@minus 0.5pt}
  {\normalfont\normalsize\bfseries}}
\makeatother

\newcommand{\sys}[0]{Atlas\xspace}

\newif\ifsubmit
\submitfalse

\ifsubmit
    \newcommand{\amir}[1]{}
    \newcommand{\sanket}[1]{}
    
    \newcommand{\todo}[1]{}
    \newcommand{\axout}[1]{}
\else
    \newcommand{\axout}[1]{\textcolor{red}{\xout{#1}}}
    \newcommand{\amir}[1]{\textcolor{magenta}{Amir: #1}}
    \newcommand{\sanket}[1]{\textcolor{cyan}{Sanket: #1}}

\fi

\newcommand{\paratitle}[1]{\noindent\textbf{\textit{#1.}}\xspace}
\newcommand{\xref}[1]{\S\ref{#1}}

\newcommand{\eg}[0]{\emph{e.g.,}\xspace}

\usepackage{cleveref} 

\crefformat{section}{\S#2#1#3} 
\crefformat{subsection}{\S#2#1#3}
\crefformat{subsubsection}{\S#2#1#3}

\newcommand*\emptycircle[1][1ex]{\tikz\draw (0,0) circle (#1);} 
\newcommand*\halfcircle[1][1ex]{%
  \begin{tikzpicture}
  \draw[fill] (0,0)-- (90:#1) arc (90:270:#1) -- cycle ;
  \draw (0,0) circle (#1);
  \end{tikzpicture}}
\newcommand*\fullcircle[1][1ex]{\tikz\fill (0,0) circle (#1);}

%% file: chapters/abstract.tex
\begin{abstract}

Cloud-mediated IoT architectures fragment authentication across vendor silos and create latency and availability bottlenecks for cross-vendor device-to-device (D2D) interactions. We present \sys, a framework that extends the Web public-key infrastructure to IoT by issuing X.509 certificates to devices via vendor-operated ACME clients and vendor-controlled DNS namespaces. Devices obtain globally verifiable identities without hardware changes and establish mutual TLS channels directly across administrative domains, decoupling runtime authentication from cloud reachability. We prototype \sys on ESP32 and Raspberry Pi, integrate it with an MQTT-based IoT stack and a \sys-aware cloud, and evaluate it in smart-home and smart-city workloads. Certificate provisioning completes in under 6\,s per device, mTLS adds only about 17\,ms of latency and modest CPU overhead, and \sys-based applications sustain low, predictable latency compared to cloud-mediated baselines. Because many major vendors already rely on ACME-compatible CAs for their web services, \sys is immediately deployable with minimal infrastructure changes.

\end{abstract}

%% file: chapters/introduction.tex
\section{Introduction}
\label{sec:introduction}

IoT ecosystems have grown rapidly, enabling programmable automation across smart homes, smart cities, and industrial environments. This programmability is largely achieved through cloud-based platforms such as IFTTT~\cite{iftttwebsite}, which allow users to compose cross-vendor device interactions using automation recipes. To simplify device management, commercial IoT platforms have adopted vertically integrated architectures in which each vendor operates its own \emph{IoT Cloud} to handle provisioning, telemetry, firmware updates, and communication. This model offers convenience and centralized control, performing adequately in single-tenant environments like smart homes where devices belong to a common administrative domain.

However, the cloud-first model introduces fundamental limitations. Cross-vendor communication requires multi-hop paths: each device must first communicate with its vendor's cloud, and a third-party service then mediates the interaction. This indirection introduces variable latency, expands the attack surface, and creates dependencies on external infrastructure. These limitations become untenable in large-scale, multi-stakeholder deployments such as smart cities, where strict quality-of-service requirements (low latency, high throughput, and continuous availability) must be maintained across independently administered device populations.

An intuitive solution is to enable direct device-to-device (D2D) communication, eliminating cloud dependency for routine interactions. However, the current IoT ecosystem lacks a standardized mechanism for secure cross-vendor authentication, an essential prerequisite for safe D2D communication. Current mechanisms typically rely on bearer tokens (\eg OAuth) mediated by cloud services, which are fundamentally application-layer constructs. This leaves the transport layer unauthenticated between devices, forcing reliance on the cloud for security. This limitation is particularly acute in multi-tenant settings where independently administered devices (\eg municipal infrastructure, emergency services, commercial systems) must securely interoperate. In the absence of a standard, deployments resort to workarounds that compromise security. Third-party platforms like IFTTT rely on shared bearer tokens that act as permanent keys, introducing significant attack surface if intercepted. Conversely, local methods such as MAC-address filtering or unauthenticated mDNS discovery provide no cryptographic identity, leaving devices vulnerable to simple spoofing attacks.

\paratitle{Fragmentation of Trust}
The root cause of this interoperability failure is \emph{trust fragmentation}. Each vendor operates as an isolated trust domain with its own certificate authority (CA) or proprietary authentication mechanism. Devices can authenticate to their vendor's cloud, but no shared trust anchor exists across vendor boundaries. A simple solution, in which vendors issue self-signed certificates, fails because it merely replicates the fragmentation problem: devices from different vendors have no basis for mutual trust. Consortium-based approaches, such as the Matter alliance~\cite{MatterCSAIoT}, attempt to address this by establishing a shared root store among member vendors. While effective for managed environments like smart homes, such alliances rely on pre-established trust relationships and manual commissioning processes (\eg scanning QR codes) that do not scale to ad-hoc, city-wide interactions. In decentralized, multi-principal deployments where municipal authorities, local businesses, and emergency services operate independently, the ``intranet'' model of alliances breaks down; an ``internet'' model of global trust is required.

\paratitle{Our Approach}
We observe that the Web faced an analogous trust fragmentation problem two decades ago, and solved it through a federated PKI model anchored in globally trusted CAs. The ACME protocol~\cite{rfc8555acme}, popularized by Let's Encrypt~\cite{letsencrypt2023stats}, further democratized this model by automating certificate issuance, renewal, and revocation. Today, ACME enables any domain owner to obtain trusted certificates at no cost, and the protocol's security properties have been formally verified~\cite{bhargavan2021depth}.

We present \sys, a framework that adapts this Web PKI model to IoT. \sys assigns each device a globally unique identity encoded as a subdomain under its vendor's domain (\eg \texttt{<uuid>.devices.vendor.com}). The vendor's IoT cloud infrastructure acts as an ACME client, performing domain validation and obtaining X.509 certificates on behalf of devices. The resulting certificates are provisioned to devices during manufacturing or via over-the-air updates. At runtime, devices authenticate directly using mutual TLS (mTLS), without requiring cloud mediation. This design preserves vendor control over device identity while enabling cross-vendor authentication through the shared trust anchor of the Web PKI.

\paratitle{Threat Model}
\sys operates under the Dolev-Yao adversary model: we assume a network attacker who can eavesdrop, intercept, modify, inject, and replay messages on the communication channel. The attacker does not have physical access to devices or the ability to compromise device firmware. We assume that vendors faithfully participate in the certificate lifecycle for the devices they administer, consistent with the Web PKI model where domain owners are trusted to assert identities within their namespace (\eg \texttt{*.vendor.com}). This trust assumption aligns with the existing IoT ecosystem, where vendors already control device firmware, credentials, and cloud infrastructure~\cite{farrukh2023one,kumar2019jedi,MCU_Token}. Device attestation systems~\cite{petzi2022scraps,hristozov2018practical,xu2019dominance}, which reason about device and firmware integrity, are complementary but outside the scope of this work.

\paratitle{Contributions}
This paper makes the following contributions:

\begin{itemize}
    \item \textbf{IoT-specific PKI framework.} We design \sys, a PKI-based authentication framework that extends the ACME protocol to IoT and provides globally verifiable identities for devices. \sys defines a DNS-based identity namespace for devices, integrates with existing vendor cloud infrastructure, and supports the full certificate lifecycle including issuance, renewal, and revocation, thereby showing that the Web PKI can serve as a practical shared trust anchor for cross-vendor IoT authentication.
    
    \item \textbf{End-to-end prototype.} We implement \sys across the IoT stack, including a vendor-side cloud component for certificate management and a lightweight device-side client compatible with resource-constrained embedded platforms. Our implementation totals approximately 2.5K lines of code, integrates with standard IoT messaging protocols, and is open-sourced to support replication and extension by other researchers.
    
    \item \textbf{Performance and scalability.} We evaluate \sys through real-world deployments on three classes of embedded hardware (Raspberry Pi 4, Raspberry Pi Zero W, ESP32) and large-scale simulations using ns-3~\cite{ns3_website}. We demonstrate that \sys provisions certificates in under 6 seconds, adds only 17ms of latency overhead per authenticated session, and scales to thousands of devices, showing that strong PKI-based cross-vendor authentication is feasible even on constrained IoT hardware. We further show that mTLS-based D2D communication achieves two to three orders of magnitude lower latency than cloud-mediated alternatives.
    
    \item \textbf{Deployment feasibility.} We analyze the practical adoption path for \sys and find that 12 of the top 20 IoT vendors already use ACME-compatible CAs~\cite{acmeclients2025} for their web infrastructure, enabling integration with minimal changes to existing deployments. We provide a security analysis under the Dolev-Yao model and discuss operational considerations including backward compatibility, revocation strategies, and privacy implications, demonstrating that \sys can be adopted within current IoT ecosystems without requiring new certification authorities or disruptive architectural changes.
\end{itemize}

%% file: chapters/background.tex
\section{Background}
\label{sec:background}

\paratitle{IoT Cloud-based Deployment}
Modern IoT ecosystems predominantly adopt a cloud-first deployment model, driven by architectural simplicity, scalability, and centralized control~\cite{aws2023iotcertificates,azure2023iotcertificates}. 

This centralized architecture simplifies development, accelerates feature rollout, and allows analytics integration without device-side changes. In practice, each vendor cloud also acts as the primary trust anchor for its devices, managing credentials and enforcing authentication policies. However, this model enforces vendor lock-in: all communication is routed through proprietary clouds, and local or cross-vendor interactions are not supported by default. For the purposes of this work, we focus on devices that utilize the IEEE 802.11 networking stack, as it is the most widely deployed wireless standard and underpins many vehicular and infrastructure-based communication systems \cite{ieee80211}. We concentrate on network-level authentication for such devices, treating device integrity mechanisms (\eg attestation) as complementary.

\paratitle{IoT Gateways and Messaging Protocols}
Despite their reliance on the Internet, IoT devices are rarely directly exposed to it. Given their constrained resources and high susceptibility to attack, these devices are typically deployed behind network gateways that provide isolation via NAT and firewalls. In smart homes, this role is usually fulfilled by a consumer-grade router, whereas in industrial and urban deployments, edge servers or aggregation points serve as IoT gateways. 
Device communication is facilitated by specialized messaging protocols such as MQTT, CoAP, XMPP, AMQP, and WebSockets that have been widely adopted for IoT, offering lightweight and reliable messaging over TCP/IP and tailored to their respective deployments. These protocols are natural termination points for authenticated channels, and later sections show how they can be secured using mutual TLS between devices.

\begin{figure}[!t]
    \centering
    \subfloat[Current cross-vendor communication approach using cloud-mediated paths.\label{fig:iot-cloud-hops}]{\includegraphics[width=0.35\linewidth]{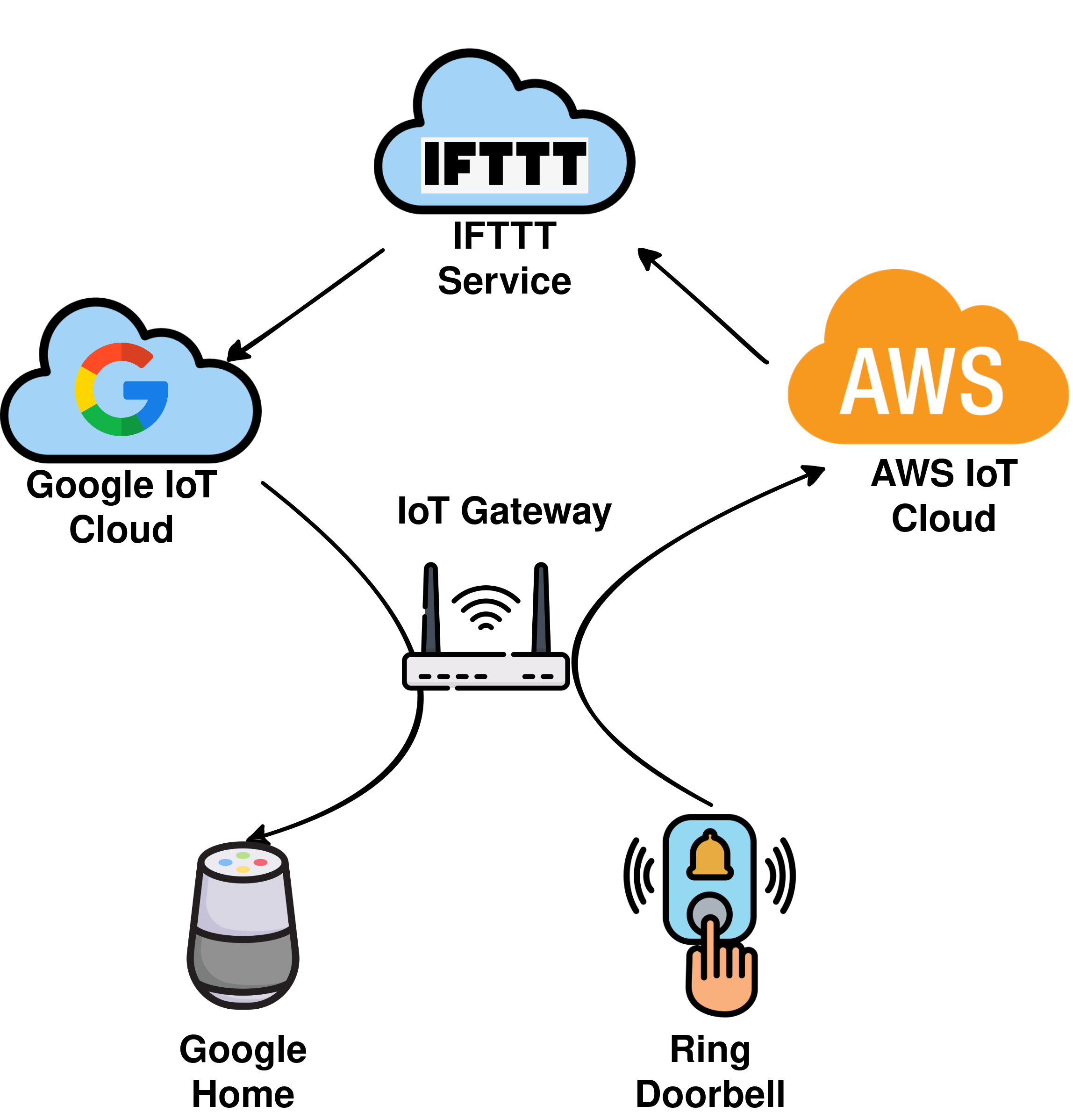}}
    \hfill 
    \subfloat[Latency distribution for cloud-mediated communication as observed in AWS IoT.\label{fig:iot-cloud-latency}]{\includegraphics[width=0.55\linewidth]{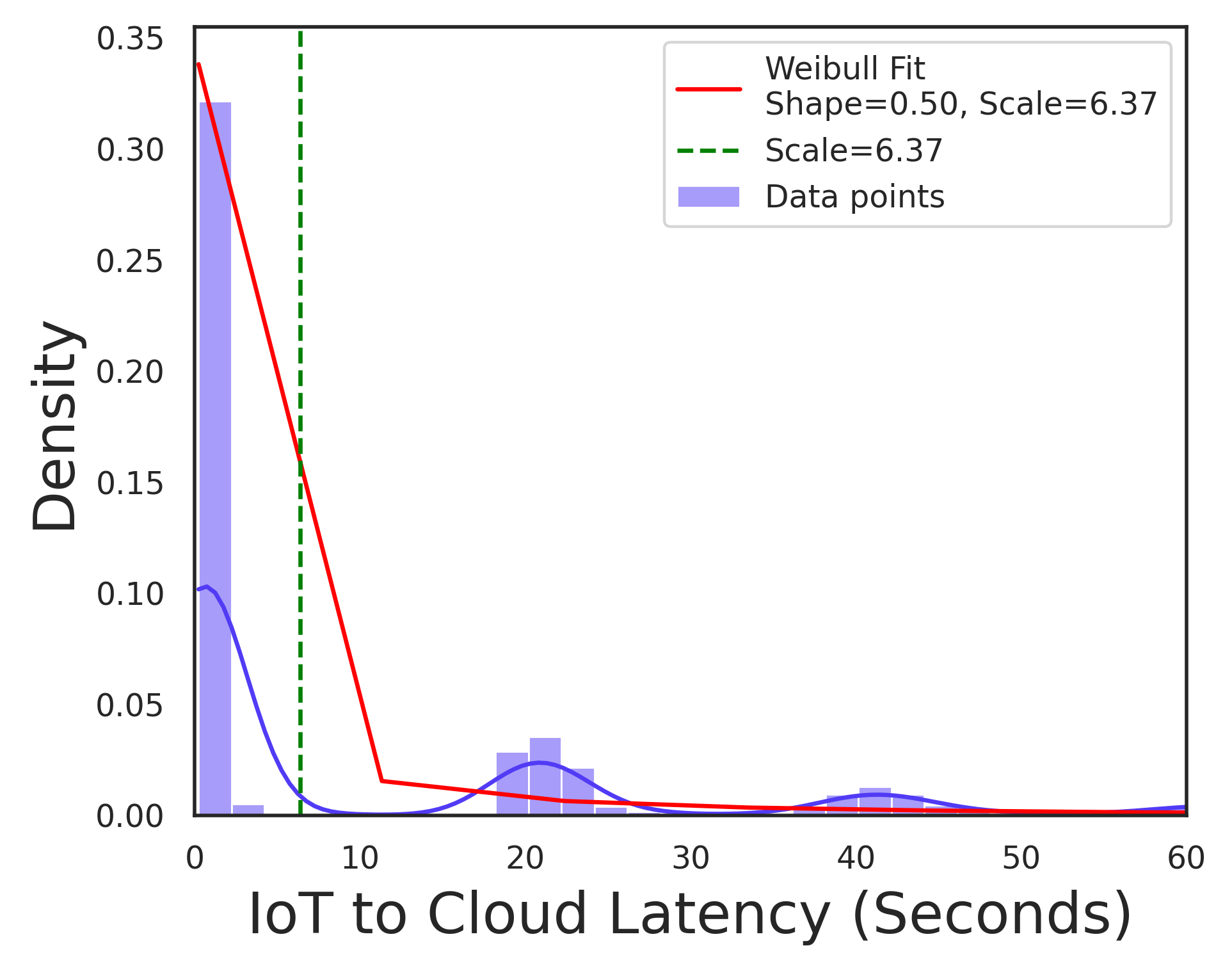}}
    \vspace{0.5em}
    \caption{Cloud-mediated IoT architectures introduce single points of failure and heavy-tailed latency (b), rendering them unsuitable for real-time cross-vendor applications.}
    \label{fig:iot-cloud}
    \vspace{-2em}
\end{figure}

\paratitle{Interoperability and Latency Costs}
The cloud-first model has the most visible impact in smart home environments, where vendors prioritize vertical integration and tight control over interoperability. Devices are hardcoded to interact exclusively with their respective IoT cloud services, resulting in a fragmented and siloed ecosystem. To enable cross-vendor functionality, third-party services act as intermediaries between isolated cloud platforms (Figure~\ref{fig:iot-cloud-hops})~\cite{iftttwebsite,zapier2025}. While these services enable limited integration, they also introduce multi-hop communication paths and complex trust dependencies, challenges that have been repeatedly highlighted in prior security research \cite{dtap,chen2021data,chen2022practical,yuan2020shattered,wang2019charting,kumar2019all,jakaria2024connectdots}. Crucially, these integrations rely on long-lived OAuth bearer tokens~\cite{yuan2020shattered,chen2021data,wang2019charting,oauth2} that act as permanent credentials; if intercepted, they grant persistent access until manually revoked. This risk is compounded by IoT protocols (\eg MQTT, CoAP) often running over unencrypted channels within local networks~\cite{alrawi2019sok}.

Beyond these well-known security concerns, cloud-mediated architectures also introduce fundamental limitations in scalability and responsiveness that have been rarely addressed in prior literature. To quantify this, we conducted a controlled experiment on AWS IoT under varying device loads (5--245 clients\footnotemark[1]) and message rates (5--100 msgs/sec) to calculate the end-to-end latency observed per client for a typical cross-vendor path (Device $\to$ AWS $\to$ IFTTT/Lambda $\to$ AWS $\to$ Device). As shown in Figure~\ref{fig:iot-cloud-latency}, the observed latency distribution follows a Weibull pattern with shape parameter $\beta = 0.5$ and scale $\lambda = 6.37$ seconds, indicating a heavy-tailed distribution with significant variance. The $\beta$ value suggests that latency becomes increasingly unpredictable as requests persist, and a non-trivial fraction of messages experience long delays (mean latency observed is $12.74$ seconds with a median of $3.06$ seconds). The high tail latency is attributable to throttling mechanisms in the cloud integration layer (AWS Lambda) during periods of high load. Although the bulk of requests complete in a few seconds, the heavy tailed distribution implies that some requests suffer disproportionately high latencies, which degrades the overall responsiveness of the system. This illustrates how cloud-mediated, multi-hop communication can be a poor fit for latency-sensitive cross-vendor interactions.

\paratitle{IoT PKI Today}
Many IoT vendors already use PKI to manage device credentials, either through in-house infrastructure or third-party providers~\cite{entrust2023,thales2023,intertrust2023}. Cloud orchestration platforms such as AWS IoT Core and Azure IoT Hub~\cite{aws2023iotcertificates,azure2023iotcertificates} expose these capabilities as managed services, and alliance-based efforts like Matter~\cite{MatterCSAIoT} define shared roots of trust for participating ecosystems. However, these deployments are vertically scoped: each ecosystem maintains its own trust anchors, and certificates issued within one ecosystem are not verifiable across others. As a result, there is no globally shared trust anchor for IoT devices that would enable seamless cross-vendor authentication. This fragmentation of PKI mirrors the interoperability and latency issues described above, and motivates our use of the Web PKI as a candidate shared trust infrastructure for IoT.

\paratitle{ACME Model}
Automated Certificate Management Environment (ACME) is a protocol designed to automate the process of domain validation and certificate issuance, renewal, and revocation~\cite{rfc8555acme}. Developed by the Internet Engineering Task Force (IETF) and popularized by Let's Encrypt, ACME eliminates manual steps in certificate lifecycle management, thereby enhancing scalability and reducing human error. The protocol has been widely adopted by global CAs deploying it at the Internet scale~\cite{acmeclients2025}. Prior to ACME, obtaining and managing a certificate required manual provisioning and proof-of-identity exchanges and often incurred financial costs, all of which discouraged widespread TLS adoption \cite{manousis2016shedding}. Today, web service providers have widely adopted the ACME model for authentication across the Internet, with Let's Encrypt alone accounting for more than $600$ million active domains \cite{letsencrypt2023stats}. These properties make ACME a natural candidate for automating certificate lifecycle management for IoT devices at scale, a direction that we explore in the design of \sys.

\footnotetext[1]{A single AWS-IoT instance seems to drop clients after 240 concurrent connections even though the developer documentations quote 500 connections per second per account is supported \cite{aws_iot_core_throttling_limits}.}

%% file: chapters/design.tex
\section{\sys}
\label{sec:system}

\sys is a PKI-based authentication framework designed to enable secure, direct communication across cross-vendor IoT devices and administrative domains. Inspired by the Web PKI model, \sys addresses the unique challenges of IoT ecosystems by defining clear design goals, tackling technical constraints, and leveraging key insights from existing authentication frameworks.

As motivated in \S\ref{sec:introduction}, we seek an authentication solution that integrates with prevalent IoT protocols, supports resource-constrained devices, and coexists with existing vendor cloud infrastructure. We adopt a network-centric threat model (Dolev-Yao): an in-path adversary can observe and manipulate traffic but lacks physical access to devices. Vendors are trusted within their namespace, consistent with the Web PKI model.

\subsection{Design Goals}
\label{sec:design-goals}
To ensure scalability and security, we identify six core principles for a cross-vendor authentication framework:

\begin{itemize}
    \item[\textbf{G1}] \textbf{Scalability Across Vendors.} Support the IoT ecosystem's growth, which includes over 16 billion devices and 14,000 vendors \cite{iotanalytic}, while maintaining efficiency across diverse deployment scenarios.
    \item[\textbf{G2}] \textbf{Cross-Vendor Interoperability.} Facilitate secure authentication between isolated vendor ecosystems through a unified trust model that respects vendor autonomy and data governance.
    \item[\textbf{G3}] \textbf{Legacy Support and Integration.} Seamlessly integrate with widely-used IoT protocols (\eg MQTT, CoAP) and existing infrastructure while ensuring backward compatibility with legacy devices.
    \item[\textbf{G4}] \textbf{Security and Revocation.} Provide robust security guarantees against adversarial threats, including impersonation and device compromise, with mechanisms for granular revocation.
    \item[\textbf{G5}] \textbf{Low Overhead.} Ensure computational efficiency and minimal communication latency, particularly for real-time systems, without compromising performance on resource-constrained devices.
    \item[\textbf{G6}] \textbf{Decentralized Trust.} Minimize reliance on centralized cloud infrastructure during runtime by prioritizing direct device-to-device authentication.
\end{itemize}

\begin{table*}[!t]
    \centering
    \caption{Comparison of current relevant approaches for establishing device identity and authentication in IoT. \\ 
    Pairing-based Encryption (PBE) establishes symmetric keys between devices through physical or contextual proximity. Identity-based Device Authentication (IDA) uses network behavior or hardware primitives to infer device identity, but \textbf{does not} support the full authentication lifecycle in the traditional sense.  Blockchain-based PKI (b-PKI) decentralizes authentication by embedding device behavior and attestation records into distributed ledgers. Traditional PKI (PKI) solutions involve widely adopted X.509 digital certificates for device authentication. See \xref{sec:related-works} for detailed analysis. \\
    \emptycircle : No Support, \halfcircle : Unlikely Adoption / Minimal Support, \fullcircle : Readily Supports, \faCheck : Currently Deployed}
    \scriptsize
    \begin{tabular}{L{0.05\linewidth}|L{0.15\linewidth}C{0.10\linewidth}C{0.10\linewidth}C{0.10\linewidth}C{0.10\linewidth}C{0.10\linewidth}C{0.10\linewidth}}
    \toprule
    \textbf{Type} & \textbf{System} & \textbf{Scalability} & \textbf{Cross-Vendor Interoperability} & \textbf{Legacy Support} & \textbf{Granular Revocation} & \textbf{Low Runtime Overhead} & \textbf{Decentralized Auth} \\
    \midrule
    \multirow{4}{*}{PBE}
    & IoTCupid~\cite{farrukh2023one} & \emptycircle & \fullcircle & \halfcircle & \fullcircle & \fullcircle & \emptycircle \\
    & UniverSense~\cite{pan2018universense} & \emptycircle & \halfcircle & \halfcircle & \emptycircle & \halfcircle & \emptycircle \\
    & T2Pair~\cite{li2020t2pair} & \emptycircle & \fullcircle & \halfcircle & \emptycircle & \halfcircle & \emptycircle \\
    & Move2Auth~\cite{zhang2017proximity} & \emptycircle & \fullcircle & \halfcircle & \emptycircle & \halfcircle & \emptycircle \\
    \cmidrule{2-8}
    \multirow{4}{*}{IDA}
    & IoT-ID~\cite{vaidya2020iot} & \emptycircle & \halfcircle & \halfcircle & \fullcircle & \halfcircle & \halfcircle \\
    & MCU-Token~\cite{MCU_Token} & \emptycircle & \halfcircle & \halfcircle & \fullcircle & \halfcircle & \halfcircle \\
    & Z-IoT~\cite{babun2020z} & \emptycircle & \emptycircle & \halfcircle & \fullcircle & \emptycircle & \emptycircle \\
    & IoT-Sentinel~\cite{miettinen2017iot} & \emptycircle & \halfcircle & \fullcircle & \fullcircle & \emptycircle & \emptycircle \\
    \cmidrule{2-8}
    \multirow{2}{*}{b-PKI}
    & Academic Proposals~\cite{neureither2020legiot,zhu2018identity,toorani2021decentralized} & \halfcircle & \fullcircle & \emptycircle & \halfcircle & \emptycircle & \fullcircle \\
    & Industry Proposal (Knox Matrix~\cite{knox-matrix}) & \halfcircle & \fullcircle & \emptycircle & \halfcircle & \emptycircle & \fullcircle \\
    \cmidrule{2-8}
    \multirow{3}{*}{PKI}
    & Vendor-controlled PKI~\cite{intertrust2023,entrust2023,thales2023} \faCheck & \emptycircle & \emptycircle & \fullcircle & \emptycircle & \fullcircle & \emptycircle \\
    & Alliance-based PKI~\cite{aws2023iotcertificates,azure2023iotcertificates,MatterCSAIoT} \faCheck & \halfcircle \footnotemark[2] & \halfcircle \footnotemark[2] & \fullcircle & \halfcircle \footnotemark[2] & \halfcircle \footnotemark[2] & \halfcircle \footnotemark[2] \\
    & \textbf{\sys [Our System]} & \fullcircle & \fullcircle & \fullcircle & \fullcircle & \fullcircle & \fullcircle \\
    \bottomrule
    \end{tabular}
    \label{tab:comparison-table}
    \vspace{-1em}
\end{table*}

Table~\ref{tab:comparison-table} compares existing approaches for IoT authentication against these design goals.

\subsection{Design Rationale: From Web PKI to IoT PKI}
The Web's PKI demonstrates how federated trust can scale across administrative domains. HTTPS servers authenticate to clients through certificate chains anchored in globally trusted root stores. This model decouples identity from single providers, supports dynamic certificate management, and scales to billions of endpoints without central coordination.

Naively applying this model to IoT is not sufficient. Devices are often deployed behind NAT gateways, lack stable public endpoints, and do not run general-purpose web servers that can participate directly in ACME-style domain validation. Vendors already terminate TLS connections and manage devices through proprietary IoT clouds, and existing IoT PKI deployments are vertically siloed, with certificates that are only meaningful within a single vendor or alliance \cite{AmazonPrivateCA2022,intertrust2023,MatterCSAIoT}. As a result, there is no globally trusted CA hierarchy for IoT devices, and cross-vendor authentication remains an open problem.

\paratitle{Key Insights}
To bridge this gap between Web PKI principles and IoT realities, \sys makes three key design choices.

\textit{First}, \sys defines a vendor-rooted DNS namespace for device identities. Each vendor controls a DNS zone (\eg \texttt{*.vendor.com}), and every device is assigned a UUID-derived subdomain within that zone. This produces globally unique, resolvable identifiers that remain under vendor administrative control, and that can be embedded directly into X.509 certificates.

\textit{Second}, \sys treats the vendor IoT cloud as the ACME client and lifecycle manager rather than the device itself. The cloud infrastructure responds to ACME challenges on behalf of devices, obtains certificates from an ACME-compatible CA, and delivers them to devices over existing management channels. This design sidesteps NAT and connectivity constraints while reusing the operational patterns that vendors already employ for firmware updates and telemetry.

\textit{Third}, \sys requires that device certificates chain to a globally shared root of trust, either in the existing Web PKI or in a future IoT-specific global PKI. This ensures that any device or gateway with the corresponding root store can validate any other device's certificate and enforce mutual TLS across vendor boundaries. Because certificates share a common trust anchor, revocation information and validation logic can also be standardized and reused across the ecosystem.

These insights guide the concrete architecture presented next. \xref{sec:system-design} describes how we instantiate these ideas using ACME and vendor clouds, and \xref{sec:design-lifecycle}--\xref{sec:design-constrained} detail how identity management, CA models, revocation, and constrained devices are handled within this framework.

\footnotetext[2]{Only supported among participating vendors. Non-participating members default to cloud-mediated setups or ad hoc solutions.}

\subsection{System Overview}
\label{sec:system-design}

\begin{figure}[!bt]
    \centering
    \includegraphics[width=0.7\linewidth]{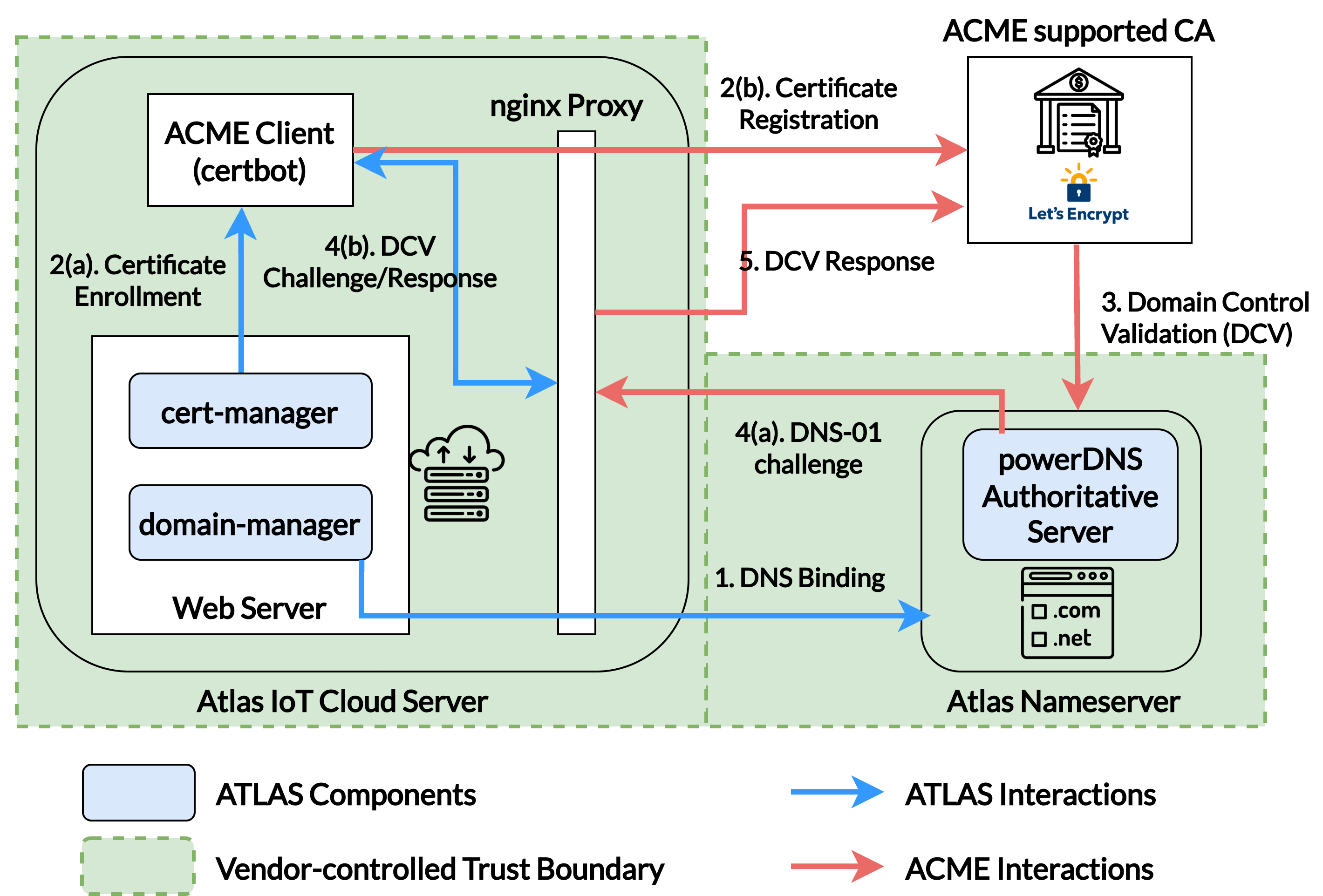}
    \caption{Deployment of \sys in vendor infrastructure: \sys integrates with existing IoT Cloud setups, leveraging ACME for certificate issuance.}
    \label{fig:system-design}
    \vspace{-2em}
\end{figure}

\sys extends the Web PKI model to IoT, allowing vendors to use existing Web CAs via ACME to issue globally verifiable X.509 certificates. As shown in Figure~\ref{fig:system-design}, \sys integrates with IoT clouds to enable cross-vendor authentication through two key stages: \textit{Domain Binding} and \textit{Certificate Issuance}.

First, each device is assigned a globally unique UUID embedded in the vendor's DNS namespace. For example, a device with UUID \textit{123a4567-b89c-12d3-a456-1234567890} from \textit{vendor.com} is assigned the subdomain \textit{<123a4567-b89c-12d3-a456-1234567890>.<vendor>.<com>}. This DNS-based identifier provides a resolvable namespace under vendor control. Second, \sys leverages ACME for certificate issuance. Vendors operate ACME challenge responders in their IoT clouds to validate device-specific domains. The CA issues X.509 certificates binding the device's DNS record to a cryptographic identity, enabling secure device-to-device and device-to-cloud authentication.

\sys ensures backward compatibility with existing IoT protocols and supports certificate revocation and renewal workflows. By aligning with the Web PKI model, \sys establishes a scalable, decentralized framework for IoT authentication.

\paratitle{Key Management and Certificate Lifecycle}
\label{sec:design-lifecycle}
\sys assigns each device a globally unique identifier within the vendor's DNS namespace and binds this identifier to a device-specific keypair and X.509 certificate. Vendors generate device keys during manufacturing, then use the \sys framework to issue the corresponding certificates from an ACME-compatible CA. Certificates are short-lived and renewed periodically through the vendor's IoT cloud, which coordinates ACME interactions and delivers updated credentials to devices over existing management channels. This lifecycle design minimizes long-term exposure of keys, supports unattended operation at scale, and allows vendors to integrate \sys with their existing provisioning and update workflows. Detailed issuance, renewal, and revocation mechanisms are described in \xref{sec:implementation}.

\paratitle{Operational CA Models}
\label{sec:design-ca-models}
\sys assumes that device certificates ultimately chain to a globally shared root of trust, such as those used in the Web PKI or a future IoT-specific global PKI. Vendors may use public ACME-compatible CAs (\eg Let's Encrypt) to obtain device certificates that are immediately verifiable by any relying party with a standard root store. Alternatively, vendors or alliances may operate ACME-based intermediate CAs that are subordinate to such global roots and issue certificates within a constrained ecosystem while still following the same protocol flows. This design ensures that mutual TLS across vendor boundaries is enforced by a common trust anchor, while still allowing flexibility in how vendors manage issuance and operational policies.

\paratitle{Revocation Strategy}
\label{sec:design-revocation}
Revocation is a critical component of any PKI-based authentication system, particularly in IoT settings where device compromise is common. \sys relies on short-lived certificates to limit the window of exposure for stale credentials and leverages Web PKI mechanisms such as CRLs and CRLite to distribute revocation information efficiently to relying parties. Because \sys issues certificates from globally recognized CAs, revocation events become standardized and globally actionable: any gateway or service that consumes the corresponding revocation feeds can immediately reject compromised devices, regardless of vendor. This stands in contrast to current IoT ecosystems, where bearer tokens and vendor-specific access keys must be manually decoupled from user accounts and third-party services, a process that prior work has shown to be error-prone and difficult to scale~\cite{dtap,yuan2020shattered}. We discuss this in detail in \xref{sec:revocation}.

\paratitle{Support for Constrained Devices}
\label{sec:design-constrained}
IoT deployments often rely on resource-constrained devices with limited CPU, memory, and storage. \sys is designed to work within these constraints by reusing existing TLS implementations on devices, relying on compact certificate chains, and supporting session resumption and hardware acceleration where available. Our experiments show that even the most resource-constrained platforms in our testbed can sustain mTLS without noticeable performance degradation, and that modern microcontrollers such as the ESP32 already integrate dedicated cryptographic accelerators specifically to support TLS workloads. As TLS libraries and hardware accelerators continue to evolve, \sys provides a standards-based path toward widespread deployment of mutual authentication in line with Zero Trust recommendations for strong, protocol-level identity verification~\cite{stafford2020zero}.

%% file: chapters/implementation.tex
\section{Implementation}
\label{sec:implementation}

This section details the implementation of \sys, structured around the IoT device lifecycle. We address four core components: \textit{Identity Creation}, \textit{Device Binding and Enrollment}, \textit{Renewal and Expiration}, and \textit{Revocation}. Together, these components demonstrate how \sys enables secure, scalable, and standards-aligned certificate management for IoT deployments.

\subsection{Identity Creation} 
\label{sec:universal-device-id}

IoT devices are typically identified by MAC addresses, which are intended to be globally unique but frequently recycled by manufacturers due to acquisition costs and production constraints~\cite{ieeeregauth}. This makes MAC addresses unsuitable as persistent global identifiers for cross-domain authentication.

\sys addresses this by using Universally Unique Identifiers (UUIDs) \cite{leach2005uuidrfc}, which require no central authority and are widely adopted for persistent identifiers \cite{microsoft-com-objects,UEFISpecWebsite,twitter-archive_snowflake}. We use version 5 UUIDs, a name-based scheme that applies a cryptographic hash over a combination of the vendor's root domain namespace and a device-bound secret \cite{rfc9562}. This produces a globally unique, reproducible identifier that is collision-resistant and independent from MAC reuse limitations. Table~\ref{table:urn-for-iot} formalizes our UUID derivation and namespace.

In a production setting, the name input to the UUID function can be derived from stronger hardware-bound secrets, such as values obtained from physically unclonable functions (PUFs) \cite{mall2022puf}, Device Identifier Composition Engine (DICE) keys~\cite{tao2021dice}, or hardware-bound keys stored in a Trusted Execution Environment (TEE). The design of \sys only requires that each device have a stable, high-entropy identifier that is unique at Internet scale and under the administrative control of its vendor. Our prototype implementation uses a synthetic identifier constructed from the MAC address and vendor domain as described above, but the framework is parametric in the choice of underlying identifier and can transparently adopt hardware-derived identities where available.

Each UUID maps to a web-resolvable subdomain under the vendor's domain, creating a structured web identity for every device. Table~\ref{table:urn-for-iot} defines a formal Uniform Resource Namespace (URN) for IoT devices of the form \(\text{UUID}.\text{device-class}.\text{root-domain}\), which can be used directly in X.509 certificates issued by any ACME-compatible CA.

\begin{table}[!t]
\centering
\footnotesize
\caption{Formal namespace for IoT devices. \emph{Each device URN embeds a UUID derived from either our synthetic UUIDv5 generator or hardware-backed keys into a vendor-controlled DNS namespace.}}
\setlength{\tabcolsep}{2pt}
\begin{bnf*}
  \bnfprod{URN}
    {\bnfpn{UUID} \bnfsp \bnfts{.} \bnfpn{device-class} \bnfsp \bnfts{.} \bnfpn{D}}\\[-2pt]
  \bnfprod{UUID}
    {\bnftd{\(\text{UUID}_{\text{synth}}\) \(\mid\) \(\text{UUID}_{\text{DICE}}\) \(\mid\) \(\text{UUID}_{\text{TEE}}\)}}\\[-2pt]
  \bnfprod{\(\text{UUID}_{\text{synth}}\)}
    {\bnftd{\(\mathrm{UUIDv5}\bigl(\mathrm{NS}_{\mathrm{DNS}}, H(D \parallel S_{\text{device}})\bigr)\)}}\\[-2pt]
  \bnfprod{\(\text{UUID}_{\text{DICE}}\)}
    {\bnftd{\(\mathrm{UUIDv5}\bigl(\mathrm{NS}_{\mathrm{DNS}}, H(D \parallel K_{\text{DICE}})\bigr)\)}}\\[-2pt]
  \bnfprod{\(\text{UUID}_{\text{TEE}}\)}
    {\bnftd{\(\mathrm{UUIDv5}\bigl(\mathrm{NS}_{\mathrm{DNS}}, H(D \parallel K_{\text{TEE}})\bigr)\)}}\\[-2pt]
  \bnfprod{device-class}
    {\bnftd{Semantic domain separator}}\\[-2pt]
  \bnfprod{D}
    {\bnftd{Vendor-controlled DNS namespace}}\\[-2pt]
  \bnfprod{\(\mathrm{NS}_{\mathrm{DNS}}\)}
    {\bnftd{UUID DNS namespace constant}}\\[-2pt]
  \bnfprod{\(S_{\text{device}}\)}
    {\bnftd{Device-bound secret}}\\[-2pt]
  \bnfprod{\(K_{\text{DICE}}\)}
    {\bnftd{Device key exposed by the DICE chain}}\\[-2pt]
  \bnfprod{\(K_{\text{TEE}}\)}
    {\bnftd{Device key stored inside the TEE}}\\[-2pt]
  \bnfprod{\(H(\cdot)\)}
    {\bnftd{Cryptographic hash function}}
\end{bnf*}
\label{table:urn-for-iot}
\vspace{-2.0em}
\end{table}

\subsection{Binding and Enrollment}
\label{sec:domaing-binding}

\begin{figure}[!tb]
    \centering
    \includegraphics[width=.95\linewidth]{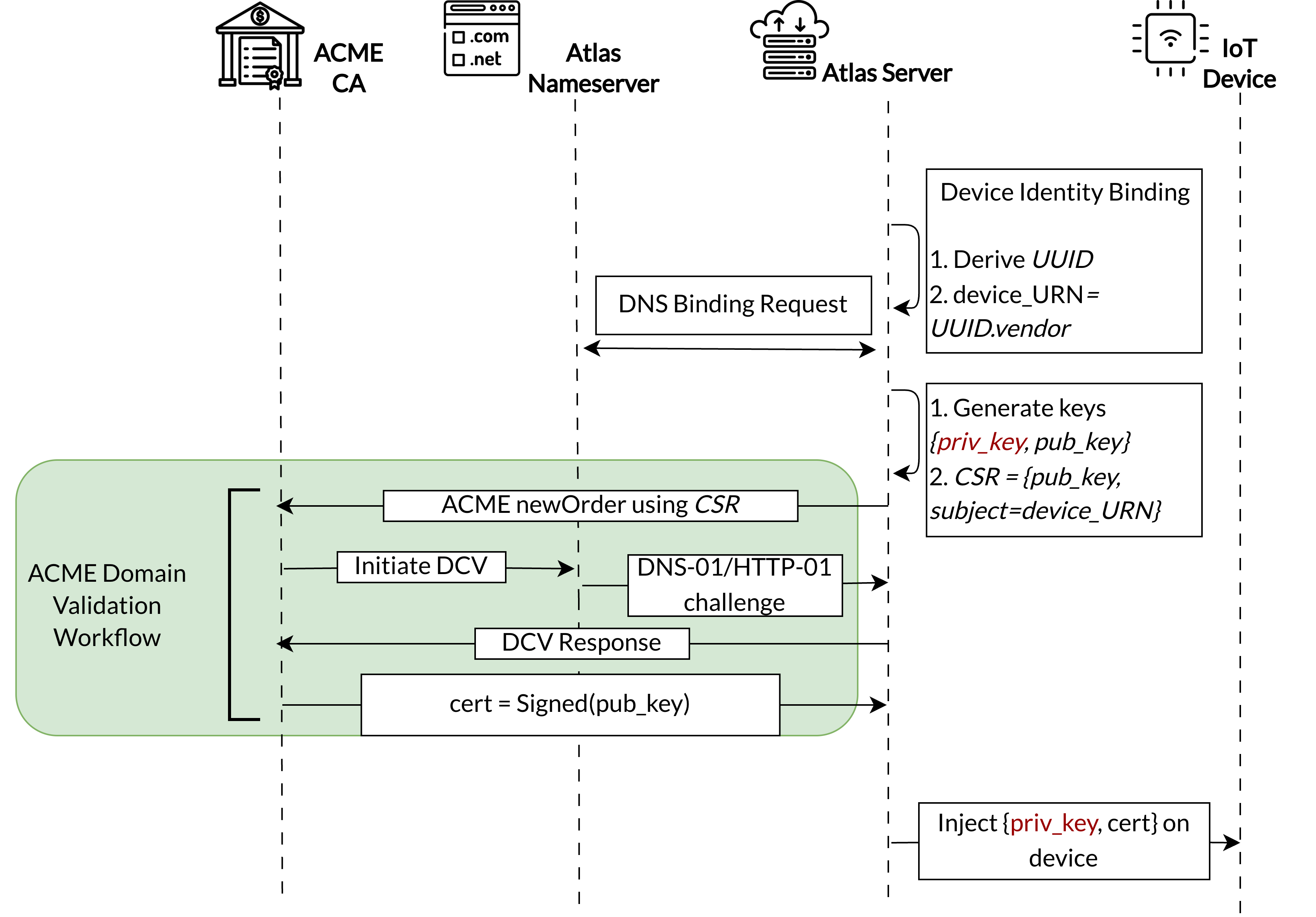}
    \caption{\emph{\sys{} Binding and Enrollment}: \sys performs DNS binding and ACME-based certificate enrollment during manufacturing. Device key pairs are generated for each device, injected once, and never stored persistently on the \sys backend; any transient copies are discarded immediately after provisioning.} 
    \label{fig:device-enrollment}
    \vspace{-2em}
\end{figure}

Unlike individually provisioned websites, IoT devices are mass-manufactured with consistent configurations. \sys extends the manufacturing workflow by provisioning each device with a certificate establishing a persistent digital identity (Figure \ref{fig:device-enrollment}).

The \sys server computes a version 5 UUID for each device using its MAC address and the vendor's root domain, generating a globally unique, deterministic identifier that mitigates MAC address reuse risks. This UUID becomes the device's canonical identifier and is stored in the \sys inventory database. The server then registers a new subdomain for the device under the vendor's domain, using the UUID as the subdomain label (\eg \textit{c69f00ac-532c-11ee-87f2-079e7eb2f068.camera.vendor.com}). This registration request goes to the \sys nameserver, an authoritative nameserver handling DNS requests for the vendor's root zone. All device subdomains use \texttt{CNAME} records resolving to the vendor's IoT cloud IP, allowing the vendor's infrastructure to terminate incoming requests without exposing devices.

\sys then initiates ACME certificate enrollment for the subdomain with any supported CA using a Certificate Signing Request (CSR). Domain control validation occurs through HTTP-01 or DNS-01 challenges executed by the vendor's ACME client. Since subdomain traffic routes to the \sys server, it can respond to ACME challenges, proving domain control and obtaining a certificate from a trusted CA. This certificate is installed on the device during provisioning. The \sys server maintains a metadata table with certificate enrollment timestamps, expiration dates, and renewal states for policy enforcement. Importantly, device private keys are ephemeral—generated during enrollment, injected into the device, then discarded from the server. This stateless approach to cryptographic material reduces key management overhead and improves security, and realizes the vendor-rooted identity and cloud-managed lifecycle design in \xref{sec:design-lifecycle}.

\subsection{Renewal and Expiration}
\label{sec:renewal}

Short-lived certificates reduce the operational risk of compromised credentials, with Let's Encrypt mandating 90-day certificates to enforce cryptographic hygiene. This is especially important in IoT ecosystems where manual intervention at scale is impractical. \sys provisions each device with a unique, domain-bound certificate from a globally trusted ACME CA that must be renewed regularly.

For devices with persistent Internet connectivity, renewals occur through an automated pipeline between the device and \sys server via the vendor's IoT cloud. The \sys Client--a lightweight, platform-agnostic agent--integrates with existing device software to handle certificate lifecycle management. Most IoT vendors already deploy heartbeat applications that periodically communicate with cloud infrastructure; \sys Client extends this pattern by embedding certificate validity checks within the same control loop, triggering renewals 30 days before expiration. The client's minimal footprint allows embedding into vendor stacks with minimal engineering overhead.

For devices with intermittent connectivity, \sys employs a proxy-based renewal mechanism. These devices typically synchronize through IoT gateways, which can mediate certificate renewals by forwarding requests to the vendor's IoT cloud. The \sys server verifies the gateway's authorization using pre-established device-gateway bindings similar to OAuth token delegations \cite{oauth2}. Throughout this process, device private keys remain secure and never transmit over the network.

Unless a compromise occurs, devices retain their original 2048-bit RSA private keys from manufacturing, conforming to CA/Browser forum requirements \cite{NIST_SP_800-57pt1r5}. This allows efficient certificate re-issuance while preserving cryptographic identity across renewals.

\begin{figure}[!tb]
    \centering
    \includegraphics[width=.95\linewidth]{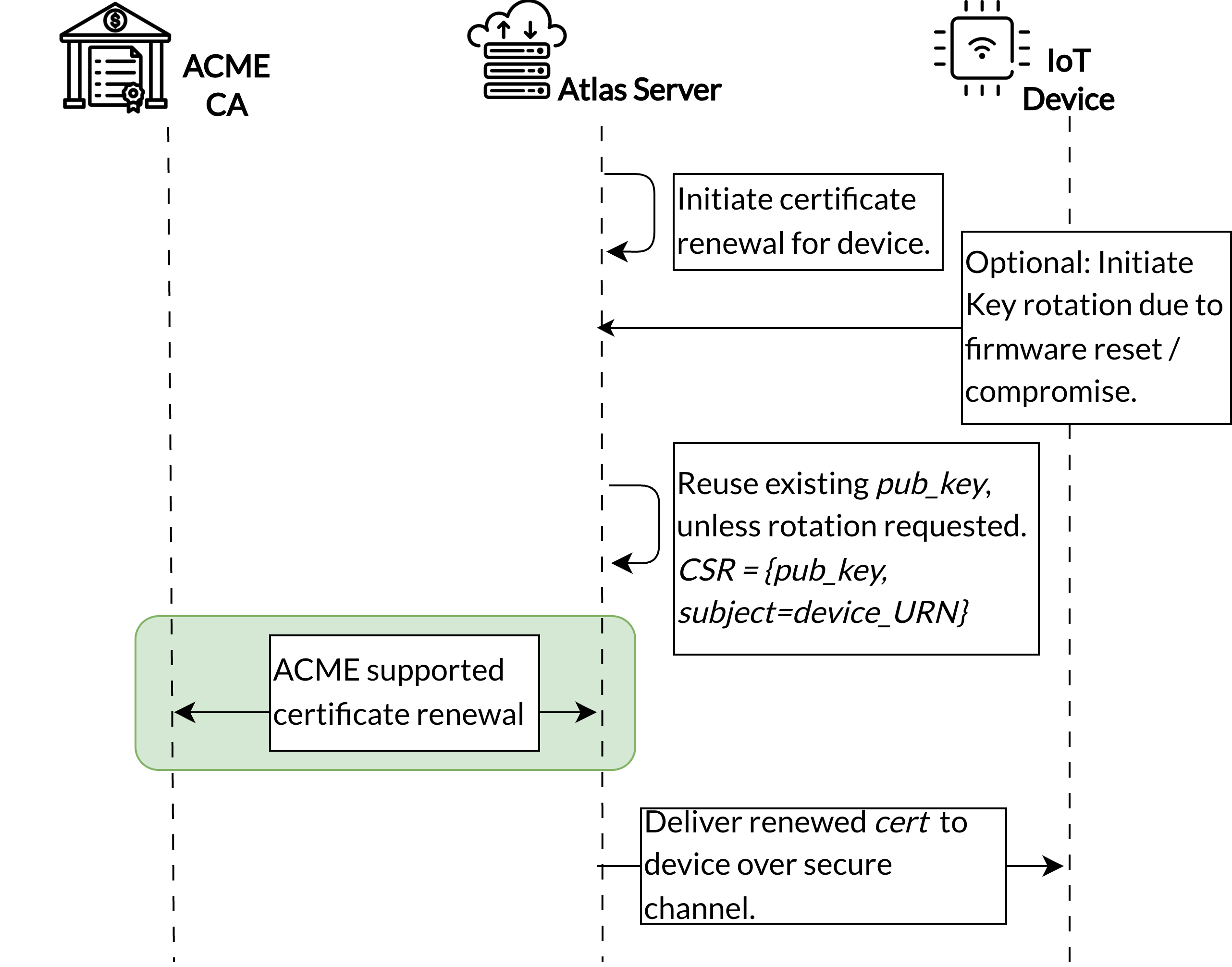}
    \caption{\sys Client runs alongside existing vendor software to perform certificate validity checks and automated renewal. Renewal requests reuse the device's existing public key while private key never leaves the device. When a new key pair must be issued (\eg after firmware refresh), the full enrollment protocol in Figure~\ref{fig:device-enrollment} is re-executed.}
    \label{fig:lifecycle-management}
    \vspace{-2em}
\end{figure}

\subsection{Revocation}
\label{sec:revocation}

IoT deployments are highly exposed to vulnerabilities, making certificate revocation critical for mitigating risks from device compromise \cite{ronen2017iot,antonakakis2017understanding}. Traditional mechanisms like Certificate Revocation Lists (CRLs) and Online Certificate Status Protocol (OCSP) are impractical for IoT due to their size, update frequency, and network requirements \cite{paracha2021iotls}. Modern browser ecosystems have adopted CRLite, which compresses revocation data into probabilistic data structures (\eg Bloom filters) that can be efficiently distributed to clients \cite{larisch2017crlite}. Our protocol follows this approach: for each vendor, the \sys backend periodically publishes a vendor-specific compressed CRL filter that encodes revoked device certificates. IoT gateways and backend services subscribe to these filters and perform local status checks during the TLS handshake, terminating connections from compromised devices without contacting the CA online. This realizes the globally actionable revocation strategy described in \xref{sec:design-revocation} while ensuring that device private keys never leave the device.

Our implementation totals approximately 2.5K lines of code across Python, C++, Arduino, and Shell scripting, and is open-sourced for future research \cite{anon2024atlas}.

%% file: chapters/evaluation.tex
\section{Evaluation}
\label{sec:evaluation}

We now evaluate \sys 
to assess its viability, performance, and comparative advantage over existing IoT authentication systems. The following research questions guide our evaluation:

\begin{itemize}[leftmargin=2em]
    \item[\textbf{RQ1}] \textbf{Device Feasibility}: Is the cross-vendor authentication model proposed by the \sys framework practical for resource-constrained IoT hardware in terms of latency and CPU overhead?
    \item[\textbf{RQ2}] \textbf{Provisioning Scalability}: What overhead does \sys introduce on vendor IoT cloud infrastructure for certificate issuance and renewal at fleet scale?
\item[\textbf{RQ3}] \textbf{Application Impact}: Can the \sys authentication framework improve the availability, resilience, end-to-end latency, and scalability of real IoT applications compared to cloud-mediated deployments?
    \end{itemize}

\paratitle{Scope} Note that, \sys{}'s evaluation in this 
section primarily focuses on feasibility and performance aspects corresponding to RQ1--RQ3 and the design goals in \xref{sec:design-goals}. We discuss its 
security analysis in the following 
section (\S\ref{sec:securityanalysis}).

\subsection{Experimental Setup}
\label{sec:experimental-setup-main}

We implement \sys across three layers of the IoT stack. \textit{Devices}: our testbed includes a Raspberry Pi 4 (RPi4)~\cite{rpi4_specs}, Raspberry Pi Zero W (RPi0W)~\cite{rpizerow_specs}, and ESP32~\cite{esp32_specs} development board, covering a spectrum from multi-core Linux-class systems to microcontrollers, all equipped with IEEE 802.11 wireless networking stack. \textit{IoT Cloud}: we deploy a \sys-compatible vendor cloud instance on a Ubuntu 20.04 server that integrates HTTPS endpoints, an application backend, an authoritative DNS server for managing per-device subdomains, and an ACME client for certificate lifecycle management (\eg certbot); for comparison, we also use AWS IoT Core as a representative cloud-mediated baseline. \textit{IoT Ecosystems}: we evaluate \sys in a local smart home testbed with an MQTT broker~\cite{mosquitto2023} and in a smart city scenario modeled in ns-3~\cite{ns3_website}, where mobile IoT nodes communicate with static gateways, and cloud-mediated latencies are emulated using an empirically derived Weibull distribution.

\subsection{Device Performance}

We quantify the impact of \sys on device-level performance and Quality of Service (QoS) by examining latency and CPU utilization across representative IoT hardware platforms under varying operational loads (\textbf{RQ1}). These metrics allow us to assess whether cross-vendor authentication via \sys (using mTLS) introduces overheads that could impair typical IoT device operations.

Our experimental design reflects the diversity of IoT messaging use cases. We focus on short-length, high-frequency communication patterns that are common in telemetry, actuation, and sensor reporting scenarios. To ensure broad applicability, we perform our experiments on three embedded systems: RPi4, RPi0W, and ESP32, reflecting the architectural capabilities of standard IoT devices. We implement a local testbed consisting of a publisher-subscriber model, where the subscriber is fixed, and the publisher is varied across the three devices. An RPi4 hosts the local MQTT broker (using Eclipse Mosquitto\cite{mosquitto2023}), emulating a typical automation hub setup. MQTT is chosen due to its lightweight nature and popularity as the de facto IoT messaging protocol.

Each publisher sends messages of 256 bytes at configurable rates (5 to 100 messages per second), holding each rate for 5 minutes. We record end-to-end latency and CPU utilization on the publisher under two configurations: (1) MQTT over TCP (insecure) and (2) MQTT over mTLS (cross-vendor authentication via \sys). This setup allows us to isolate the cost of secure channel establishment and sustained cryptographic processing required for applications using the \sys framework. For the ESP32, we implemented MQTT over TLS instead of full mutual TLS due to a known limitation in the underlying mbedTLS library \cite{mbedtls_issue_6566}. However, we note that mTLS is simply TLS with additional client certificate verification done by the server, establishing a functionally equivalent secure channel for the purposes of this evaluation.

\begin{figure*}[!tb]
    \centering
    \includegraphics[width=0.75\textwidth]{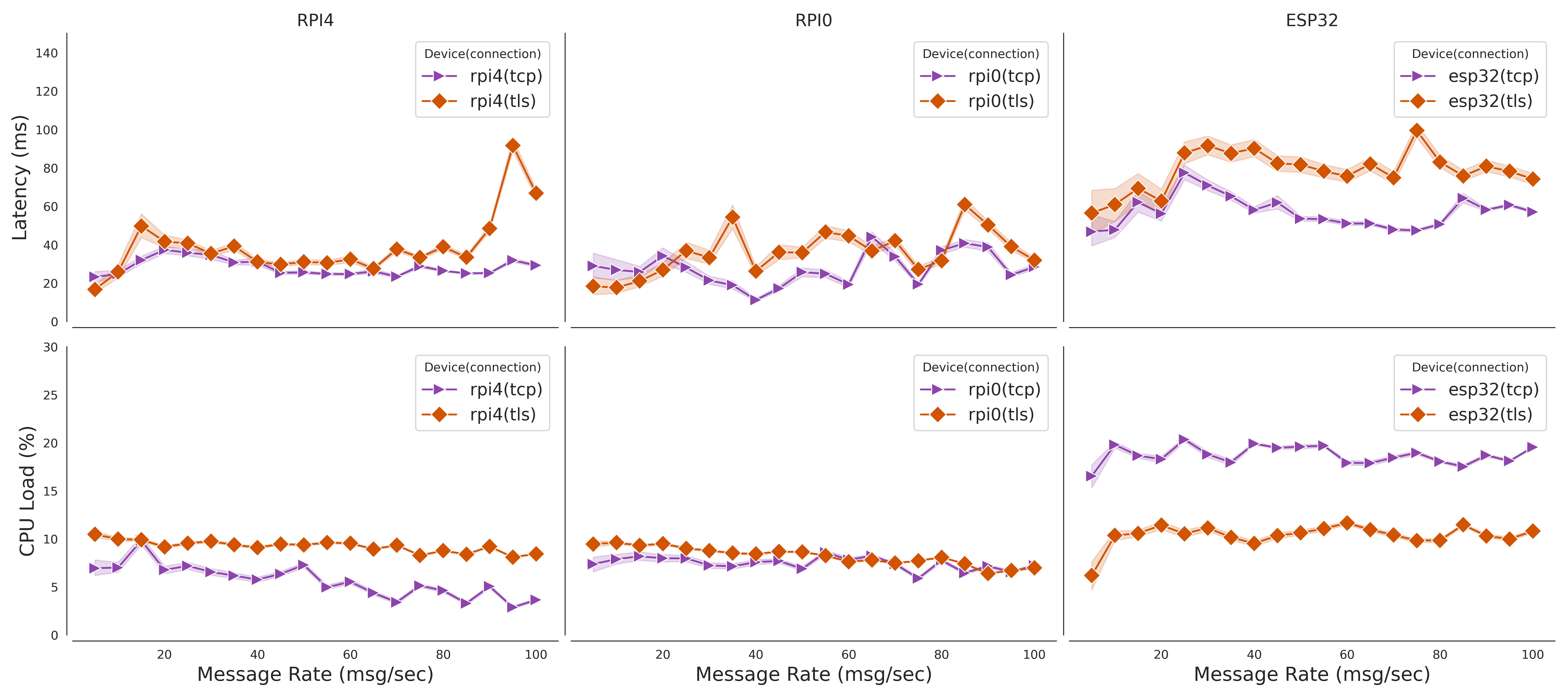}
    \caption{Impact of mTLS on Resource-Constrained Devices: \sys introduces negligible latency ($\sim$17ms) and CPU overhead (<9\%), demonstrating feasibility even on microcontrollers (ESP32).}
    \label{fig:device-performance}
    \vspace{-1em}
\end{figure*}

\noindent\textbf{Findings:}
Figure~\ref{fig:device-performance} compares latency and CPU utilization across varying message rates for each platform. Table~\ref{tab:device-overhead} presents a consolidated summary, reporting mean end-to-end latency and average CPU load for each device under both insecure (MQTT over TCP) and secure (MQTT over mTLS) configurations.

\noindent\textbf{Takeaways:}
\begin{itemize}
    \item \textit{Low Latency Overhead:} Secure channel establishment via TLS or mTLS introduces minimal latency overhead. On average, secure communication adds 16ms on the RPi4, 10ms on the RPi0W, and 24ms on the ESP32 (see Table~\ref{tab:device-overhead}). These delays, attributed to the TLS handshake, asymmetric key exchange, and additional payload padding, remain within acceptable bounds for any latency-sensitive application. Overall, MQTT over TLS imposes only $\sim17ms$ additional latency across devices---well within the real-time performance thresholds of many IoT applications.
    
    \item \textit{Negligible CPU Impact on Low-Power Devices:} TLS introduces only a marginal increase in CPU utilization. On the RPi4, secure communication results in a $4.14\%$ increase in CPU load, while on the RPi0W, the added load is just $0.4\%$. This efficiency stems from lightweight TLS libraries, which are optimized for embedded platforms. For instance, session reuse and caching mechanisms mitigate repeated handshake overheads for frequent connections, improving CPU efficiency during sustained device usage.

    \item \textit{Hardware Acceleration in Modern Boards:} We draw attention to the CPU performance on ESP32 (Table~\ref{tab:device-overhead}), where counter-intuitively \textit{TLS is $8\%$ faster than TCP}. This is because many modern microcontrollers are equipped with onboard co-processors for cryptographic hardware acceleration. mbedTLS library offloads all TLS handshake operations to this dedicated hardware, minimizing load on the general-purpose CPU, enhancing application performance directly \cite{wolfssl_espressif_readme}.

\end{itemize}

\begin{table}[!tb]
\centering
\caption{Latency measurements show that the overhead of TLS is well within the real-time performance thresholds for IoT. Modern micro-controllers (\eg esp32) have dedicated cryptographic accelerators that offload TLS computation.}
\small
\setlength{\tabcolsep}{3pt}
\begin{tabular}{@{}l|lccc@{}}
\hline
\multicolumn{2}{@{}l@{}}{\textbf{Metric}} & \textbf{rpi4} & \textbf{rpi0} & \textbf{esp32} \\
\hline
\multirow{2}{*}{\textbf{Latency (ms)}} & tcp & 27.68 & 29.25 & 56.54 \\
\cline{2-5}
& tls & 44.28 & 39.47 & 80.76 \\
\hline
\multirow{2}{*}{\textbf{CPU (\%)}} & tcp & 4.80 & 7.28 & 18.69 \\
\cline{2-5}
& tls & 8.94 & 7.69 & 10.56 \\
\hline
\end{tabular}
\label{tab:device-overhead}
\vspace{-1em}
\end{table}

\subsection{IoT Cloud Overhead}

For \textbf{RQ2}, we focus on quantifying the latency introduced by \sys during device provisioning at scale, particularly evaluating its implications for certificate lifecycle management. IoT vendors require enrollment processes that are fast, scalable, and automated, as manual or slow provisioning directly hinders manufacturing throughput and product deployment timelines. Using our experimental IoT cloud setup (Figure~\ref{fig:system-design}), we measure the latency of the two critical operations in \sys provisioning: (1) Domain Binding, where a unique device identity is associated with a DNS namespace under the vendor's zone, and (2) Certificate Issuance, where \sys invokes the ACME protocol to validate domain ownership and retrieve a device certificate. We use Let’s Encrypt as the ACME-compatible CA, which operates ACME at the Internet scale and supports short-lived certificate issuance. To emulate bulk provisioning scenarios, we conducted enrollment batches ranging from 50 to 500 devices. \textit{Domain Binding} operations were performed using our authoritative DNS server, while \textit{Certificate Issuance} was executed against the Let’s Encrypt ACME staging endpoint~\cite{letsencrypt_staging}. The staging CA mirrors Let's Encrypt's production setup and enforces equivalent rate limits---allowing up to $30,000$ certificate requests per week per registered domain.

\noindent\textbf{Findings and Takeaway:}
The \sys framework incurs an average of $0.03\pm0.008$ seconds per device for Domain Binding and $4.85\pm1.36$ seconds per device for Certificate Issuance. These costs are incurred primarily at manufacturing time or during scheduled renewals and demonstrate that ACME-based provisioning is fast enough to support fleet-scale deployments without becoming a bottleneck for vendors.


\subsection{IoT Applications Using \sys Framework}
In order to quantify the performance benefits of using \sys in real IoT applications (\textbf{RQ3}), we design and execute two sets of experiments using the smart home and smart city setups described in \xref{sec:experimental-setup-main}.

\paratitle{Experiment 1: Smart Home Automation}
We begin with our smart home testbed to evaluate the end-to-end latency in typical IoT messaging scenarios. A local Mosquitto MQTT broker is used to implement a trigger-action automation system within a LAN environment. For comparison, we replicate the same functionality using AWS IoT infrastructure, where Raspberry Pi 4 (RPi4) devices, provisioned with the AWS IoT SDK, act as publishers and subscribers. The cloud-hosted trigger-action logic is implemented using AWS Lambda. This baseline mirrors the current IoT cloud architecture, as previously illustrated in Figure~\ref{fig:iot-cloud-hops}. Our measurements directly compare the two deployment paradigms: (1) secure, direct D2D applications using \sys, and (2) traditional cloud-mediated applications through AWS IoT.

\paratitle{Experiment 2: Smart City Automation}
To evaluate scalability, we utilize our smart city simulation to model a representative use case involving Internet-connected vehicular infrastructure. Large-scale real-world testbeds are typically infeasible; thus, we employ the ns-3 discrete-event simulator to emulate a smart city network. The simulation consists of mobile IoT devices (\eg autonomous vehicles, public transit systems) that interact with static IoT gateways (\eg traffic management servers).

We simulate a variable number of mobile nodes (ranging from 50 to 2500) and maintain a fixed set of five static IoT gateways positioned in a star topology centered around the city core to emulate realistic urban network coverage. Vehicular movement is modeled using the widely adopted \textit{RandomWaypointMobilityModel} from the ns-3 library. Each mobile IoT device runs a client application that sends periodic messages to its nearest gateway, which hosts a server-side application. This client-server model is used to quantify the benefits of direct device-to-device (D2D) communication enabled by \sys.

To accurately model the latency induced by IoT cloud mediation, we introduce a synthetic delay modeled using a Weibull distribution (Figure~\ref{fig:weibull-formula}) with parameters $\beta = 0.5$ and $\lambda = 6.37s$. These parameters are empirically derived from latency measurements observed in AWS IoT setups used in \textit{Experiment 1}. The empirical latency distribution is presented in Figure~\ref{fig:iot-cloud-latency}.

\begin{figure}[tb]
    \centering
    \footnotesize
    \[
    f(x; \beta, \lambda) = \frac{\beta}{\lambda} \left( \frac{x}{\lambda} \right)^{\beta-1} \exp\left( -\left( \frac{x}{\lambda} \right)^\beta \right) \quad \text{for } x \ge 0
    \]
    \caption{Weibull distribution used to model cloud-mediated latency, parameterized by $(\beta,\lambda) = (0.5, 6.37\,\text{s})$ as derived from AWS IoT measurements.}
    \label{fig:weibull-formula}
    \vspace{-2em}
\end{figure}

\begin{figure}[tb]
    \centering
    \subfloat[Smart home workload (RPi4).]
    {\includegraphics[width=0.48\linewidth]{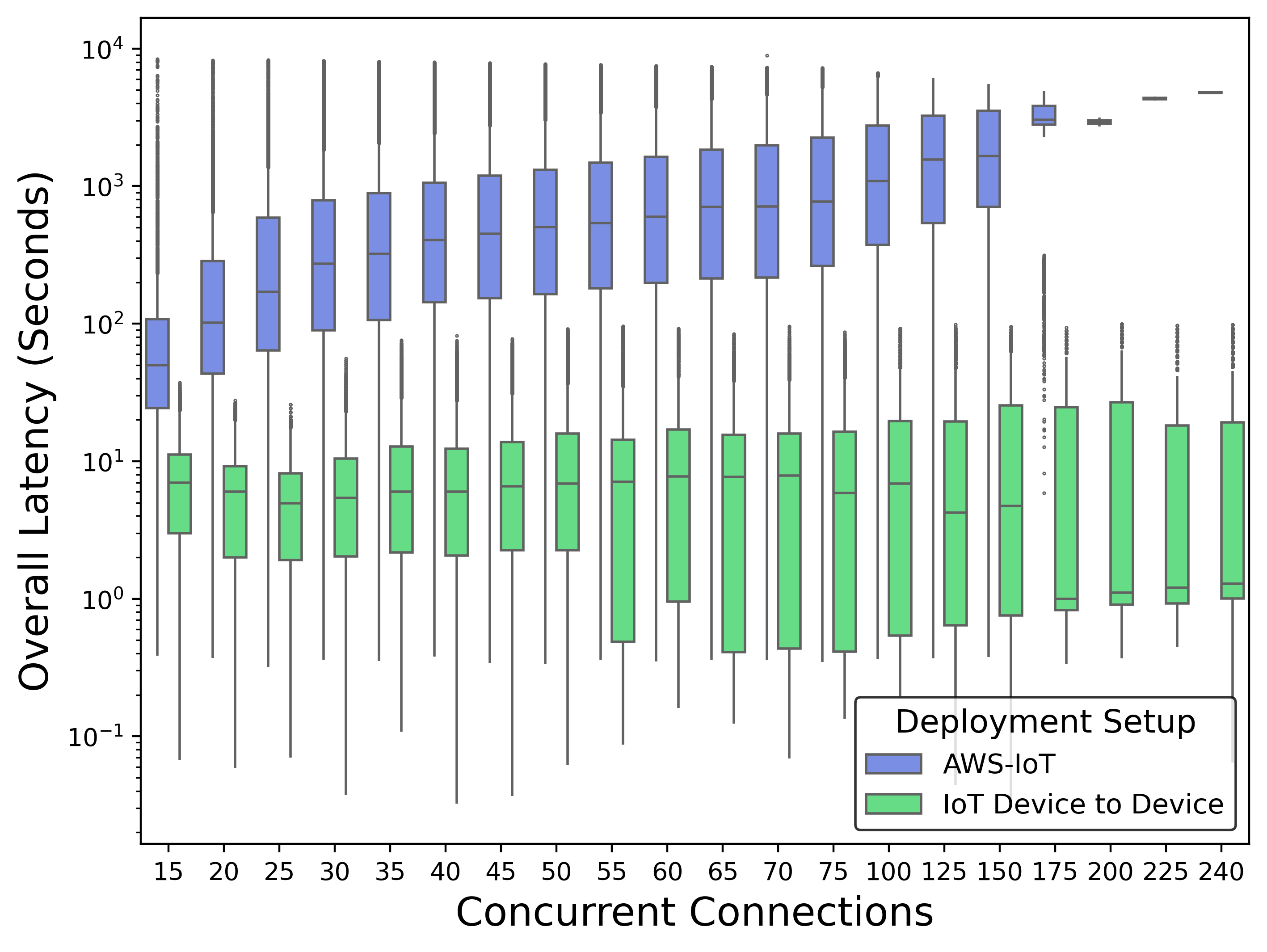}
    \label{fig:concurrent-load}}
    \hfill
    \subfloat[Smart city workload (ns-3 simulation).]
    {\includegraphics[width=0.48\linewidth]{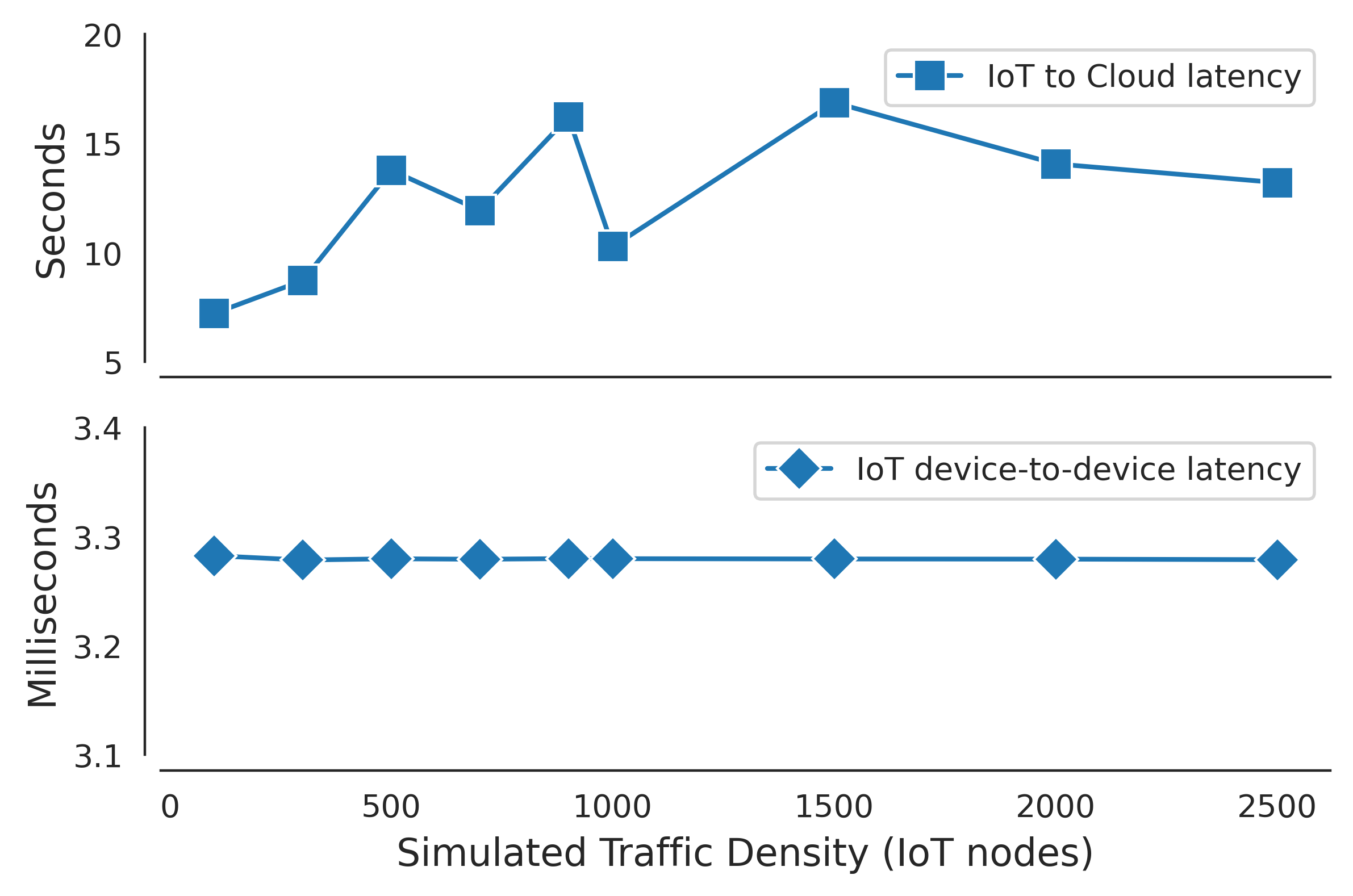}
    \label{fig:simulation-latency}}
    \caption{Performance comparison under load. (a) Smart Home: \sys maintains stable, low latency while AWS IoT exhibits high variance and packet drops. (b) Smart City: \sys scales to thousands of nodes with millisecond-level latency, contrasting with the heavy-tailed, multi-second delays of cloud mediation.}
    \label{fig:combined-performance}
    \vspace{-2em}
\end{figure}

\noindent\textbf{Findings:}
Figure~\ref{fig:concurrent-load} demonstrates the performance of smart home IoT applications using \sys framework under a scaled workload, while Figure~\ref{fig:simulation-latency} illustrates the latency distribution observed in large-scale smart city-style deployments powered by \sys versus a simulated cloud-mediated baseline.

\noindent\textbf{Takeaway:}
Our results indicate that traditional IoT cloud infrastructures are only feasible in limited contexts: primarily single-tenant, latency-tolerant smart home environments with low message throughput and tightly coupled vendor ecosystems. However, as communication demands scale---either through higher message rates or increased device count---cloud-mediated architectures exhibit significant performance degradation. In \textit{Experiment 1}, the AWS IoT setup displayed substantial latency variance under moderate load and dropped most packets beyond $150$ concurrent connections, even though developer documentations claim that AWS-IoT can support higher throughput (upto $500$ concurrent connections~\cite{aws_iot_core_throttling_limits}). In \textit{Experiment 2}, cloud-mediated communication regularly exceeded 10 seconds of delay per device, even for basic message exchanges. In contrast, IoT applications using \sys demonstrate consistent, low-latency communication making it suitable for real-time and safety-critical systems.

\subsection{Limitations and Threats to Validity}

Our evaluation is subject to several limitations. First, we do not empirically study availability under CA or cloud outages; our measurements capture performance in steady-state operation, and we leave fault-injection studies to future work. Second, while we analytically discuss CT privacy and CA rate limits in the design and security analysis, we do not stress-test these dimensions experimentally. Third, our ESP32 experiments use TLS rather than full mTLS due to library limitations; however, since the client still performs certificate validation and the underlying cryptographic workloads are comparable, we expect the latency and CPU overhead conclusions to generalize to mTLS-capable firmware.

\section{Qualitative Security Analysis of \sys}
\label{sec:securityanalysis}

\subsection{\sys End-to-End Security}
We analyze the end-to-end security of the \sys framework under the Dolev-Yao threat model (Section \ref{sec:introduction}), wherein the adversary can intercept, modify, inject, replay, or drop any network message. Our analysis considers all critical components of \sys : IoT Cloud, ACME CA, and IoT Device, and evaluates how the design is robust against the canonical classes of Dolev-Yao attacks.

\paratitle{Trusted Infrastructure}
The IoT Cloud includes the vendor-operated web server and authoritative nameserver, both within the vendor's trust boundary and protected via standard practices (firewall enforcement, TLS-only communication, intrusion prevention via fail2ban~\cite{Fail2Ban}). Internal communications use static IP routing, eliminating DNS-based resolution within vendor infrastructure and preventing MitM and DNS poisoning attacks. \sys deploys PowerDNS~\cite{PowerDNS} as its authoritative nameserver, enabling real-time subdomain issuance and granular TTL configuration without registrar rate limits or propagation delays. The DNS zone is tightly controlled and accessible only by authenticated administrators.

\paratitle{CA Interactions}
Certificate issuance in \sys is handled using the ACME protocol, which has been formally verified for security~\cite{bhargavan2021depth}. The client (\eg Let's Encrypt's certbot) interacts with the CA over HTTPS for control messages including certificate requests, challenge submission, and certificate downloads. These HTTPS interactions ensure protection against MitM, impersonation, and downgrade attacks. Domain validation itself (\eg via http-01) is performed over unauthenticated HTTP so that the CA can validate actual control over the domain by probing a specific token served by the device or cloud server. While HTTP is used for this validation step, the challenge itself is cryptographically bound to the client’s ACME account key, and includes fresh, signed tokens. This mechanism prevents spoofing, replay, and reflection attacks. Only clients who can both respond to the challenge and prove possession of the private key can successfully obtain a certificate.

\paratitle{Device Certificate Provisioning}
Each IoT device is provisioned during manufacturing with a unique X.509 certificate and keypair. This step occurs under vendor control and ensures that each device has a globally unique and verifiable cryptographic identity. Certificate renewals are handled periodically via over-the-air (OTA) updates, which are conducted over TLS channels and authenticated by both client and server using mutual TLS (mTLS). This setup ensures that all subsequent device communications are cryptographically authenticated and encrypted, providing protection against eavesdropping, impersonation, injection, and message tampering. TLS session freshness and challenge-response authentication also prevent replay attacks. The use of TLS 1.3 and controlled cipher suite negotiation ensures that protocol downgrade attacks are not feasible.

\subsection{Additional Privacy and Security Considerations}

\paratitle{Privacy Issues}
Device subdomains appearing in Certificate Transparency (CT) logs may expose metadata about a vendor’s deployment or device inventory, raising competitive privacy concerns. However, \sys provides multiple mitigations. First, CT entries do not indicate whether a certificate is ever used or deployed, allowing vendors to generate excess certificates as decoys. Second, device URNs are constructed to allow anonymized identifiers, such as UUIDs, without semantic information about device types or capabilities. Finally, vendors with stricter privacy requirements can operate their own ACME-compatible intermediate CA, bypassing public CT altogether while remaining compliant with \sys’s operational model.

\paratitle{Malicious Vendor}
\sys assumes that vendors will faithfully participate in the certificate lifecycle, including proper key generation, renewal, and revocation. A malicious or negligent vendor could, in theory, retain copies of device private keys or ignore renewal requirements, undermining security guarantees. However, this threat is not unique to \sys; all modern consumer devices---including smartphones and operating systems---rely on vendors for essential security updates. Moreover, a vendor already has complete control over its devices in vertically integrated systems, so \sys does not introduce new incentives or opportunities for abuse. Rather, it provides a structured mechanism for vendors to manage device credentials more securely at scale.

\paratitle{Domain Impersonation}
A legitimate concern is whether adversaries can abuse \sys’s domain-based identity model via techniques like typosquatting~\cite{szurdi2014long}, homograph attacks~\cite{roberts2019you}, or acquiring expired vendor domains~\cite{so2022domains}. \sys explicitly mitigates these risks by issuing certificates only for subdomains that are cryptographically bound to an active, validated parent domain controlled by the vendor. The use of ACME ensures that domain validation is enforced before certificate issuance. Additionally, device identities are embedded in structured URN namespaces derived from UUIDs and bound to the vendor’s domain, making them cryptographically unique and not forgeable without domain control. Even if certificate transparency (CT) logs expose device subdomains, those entries always resolve to vendor-managed infrastructure, never exposing the device directly.

\paratitle{Compromised Private Key} 
IoT devices are known to be vulnerable due to their constrained resources and limited physical protections, making them frequent targets for compromise~\cite{ronen2017iot, antonakakis2017understanding}. A critical failure scenario is the leakage of a device’s private key, which enables impersonation attacks against systems that implicitly trust the device. Unlike many real-world IoT deployments where certificate reuse and hardcoded credentials are still prevalent~\cite{kilgallin2019factoring}, \sys confines the scope of such compromise: every device in \sys has a unique private key and certificate, ensuring that a single compromised device does not affect the broader ecosystem. While \sys facilitates post-compromise mitigation via certificate revocation (\cref{sec:revocation}), it does not address the detection or prevention of such compromises. These challenges are orthogonal to \sys and can be addressed by complementary runtime attestation or behavioral monitoring solutions~\cite{miettinen2017iot,oconnor2019homesnitch}.

%% file: chapters/discussion.tex
\section{Discussion}
\label{sec:discussion}


\paratitle{Backwards Compatibility}
Vendors who currently use proprietary provisioning workflows can integrate \sys without overhauling their backend systems. Since certificate installation occurs on-device, existing devices can be updated to adopt \sys-based authentication through over-the-air (OTA) firmware updates. This allows a graceful migration path for existing fleets while preserving compatibility with legacy device management tools.

\paratitle{Incentive Compatibility}
A key motivation for adopting ACME in \sys is the economic and operational efficiency it brings. Many ACME-supported CAs already offer zero-cost certificate issuance \cite{acmeclients2025}. IoT vendors can readily benefit from streamlined, zero-cost certificate issuance that scales effortlessly with device production. This incentive alignment of zero cost, higher automation, and formal-verification-backed stronger security positions \sys as a practical alternative to fragmented, proprietary provisioning systems that dominate today’s IoT landscape.

\paratitle{IoT-specific PKI or not}
A common concern with \sys is whether ACME-based global PKI infrastructures can scale to meet the combined needs of the Web and IoT. Our evaluation suggests that existing ACME deployments are already capable of supporting dual use for web services and IoT devices, given their robust infrastructure and automation support. At the same time, \sys is intentionally agnostic to the choice of CA: vendors can rely on public Web CAs, operate IoT-specific intermediate CAs beneath a shared root, or deploy private ACME-compatible CAs within alliances. As ACME adoption grows, we expect additional CAs and intermediate hierarchies to emerge, providing tailored support for IoT workloads while preserving the globally verifiable trust roots that \sys relies on~\cite{acmeclients2025}.

%% file: chapters/related_works.tex
\section{Related Work}
\label{sec:related-works}

Existing IoT authentication frameworks span four broad categories---device pairing, identity derivation, blockchain-based PKI, and traditional PKI-based provisioning. While each addresses specific challenges, none simultaneously achieves the scalability, cross-vendor interoperability, lifecycle management, and decentralized trust captured by the design goals in \xref{sec:design-goals} and summarized in Table~\ref{tab:comparison-table}.

\paratitle{Pairing-based Encryption (PBE)}
Pairing-based schemes establish symmetric keys between devices via human or contextual interaction (\eg barcodes, gestures, co-location)~\cite{zhang2017proximity,li2020t2pair,farrukh2023one}. They are well-suited to small personal deployments but do not scale to thousands of devices, since each association requires physical proximity or user effort. In contrast, \sys moves identity establishment into the network fabric (DNS and PKI), enabling unattended, factory-scale provisioning and cross-vendor authentication (G1, G2, G3).

\paratitle{Identity-based Device Authentication (IDA)}
Behavioral and fingerprinting approaches such as IoT-Sentinel, IoTFinder, Z-IoT, and PUF-based schemes~\cite{miettinen2017iot,perdisci2020iotfinder,babun2020z,mall2022puf,MCU_Token} infer or derive identifiers from traffic patterns or hardware characteristics. These are effective for classification and anomaly detection but do not provide stable, verifiable identities with standard revocation and re-keying semantics. \sys complements such techniques by binding devices to globally verifiable X.509 identities and a full certificate lifecycle (G3, G4).

\paratitle{Blockchain-based PKI (b-PKI)}
Blockchain-based proposals such as SerenIoT, LegIoT, and Knox Matrix~\cite{thomasset2020sereniot,neureither2020legiot,knox-matrix} offer decentralized trust and auditability by recording behavior and policy in distributed ledgers. However, they introduce significant synchronization and consensus overheads and have seen limited real-world deployment in resource-constrained IoT. \sys instead relies on the existing federated trust of DNS and ACME, achieving interoperability and global revocation (G1, G2, G4, G6) without requiring new consensus infrastructure.

\paratitle{Traditional and Cloud-Hosted PKI (PKI)}
Commercial PKI offerings and IoT cloud platforms~\cite{entrust2023,thales2023,intertrust2023,aws2023iotcertificates,azure2023iotcertificates,MatterCSAIoT} provide strong per-vendor or per-alliance authentication, but remain vertically siloed: trust roots are limited to a single provider set and cross-vendor interactions typically fall back to cloud mediation or ad hoc protocols. \sys generalizes these efforts by rooting device identities in vendor-controlled DNS namespaces under a globally shared CA hierarchy, so that any compliant gateway or service can validate certificates across vendors while vendors retain control over their own namespaces and lifecycle policies (G2, G4, G6).

%% file: chapters/conclusion.tex
\section{Conclusion}
\label{sec:conclusion}

In this paper, we presented \sys, a unified authentication framework that enables secure, cross-vendor device-to-device communication by extending ACME and globally shared CAs to vendor-controlled DNS namespaces for IoT devices. \sys supports the full certificate lifecycle, including issuance, renewal, and revocation, without requiring fundamental architectural changes from vendors, while enabling mutual TLS between heterogeneous devices under a network-centric threat model. Our implementation and evaluation on constrained hardware, a \sys-compatible cloud stack, and smart home and smart city workloads show that \sys is feasible and scalable in practice, delivering mTLS-based authentication with modest overhead and significantly lower, more predictable latency than cloud-mediated baselines; by enabling direct D2D communication, \sys can improve service availability during Internet or cloud disruptions by avoiding dependence on continuous cloud mediation.


%% file: chapters/appendix.tex
\section{\sys Certificate Lifecycle}
\label{sec:cert-lifecycle}

\subsection{Validation}
\begin{figure}[!tb]
    \centering
    \includegraphics[width=.6\linewidth]{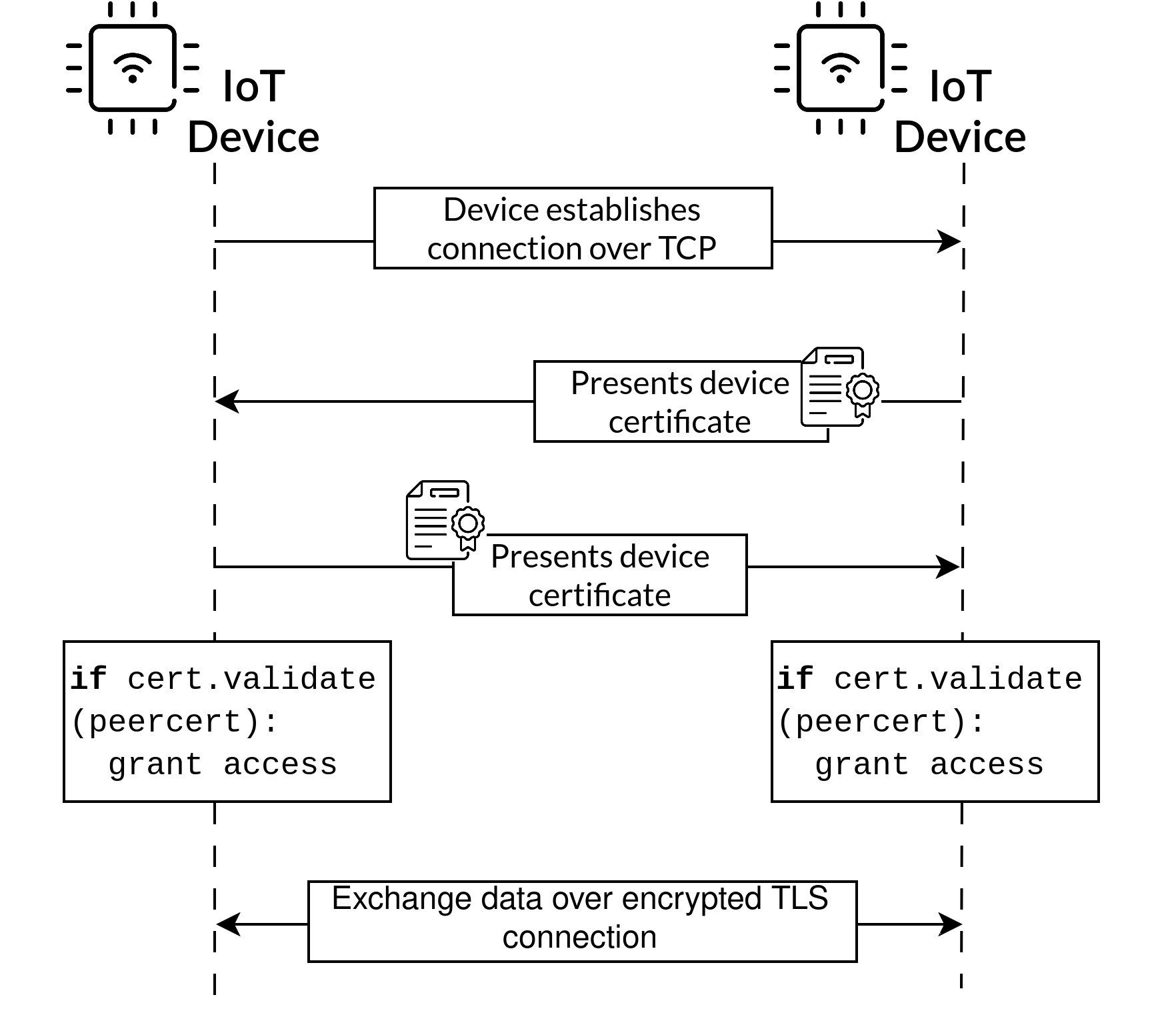}
    \caption{Cross-Vendor Authentication using \sys: \sys provides a model for IoT applications to adopt cross-vendor D2D authentication using mTLS.}
    \label{fig:validation}
    \vspace{-1em}
\end{figure}

\sys provisions each IoT device with a globally valid, device-specific certificate from a publicly trusted CA, enabling direct mutual authentication via mTLS without cloud mediation. As shown in Figure~\ref{fig:validation}, two devices—potentially from different vendors—can securely establish end-to-end channels within a local network. 

Unlike traditional cloud-mediated setups that introduce latency and centralize trust, \sys leverages the shared trust anchor of the Web PKI to support cross-vendor interoperability. Since all device certificates descend from a public CA, devices can independently validate each other's credentials. The mTLS handshake uses standard TLS libraries (\eg OpenSSL, mbedTLS), which verify the peer’s certificate chain, validity, and revocation status against system-trusted roots. Certificates issued by \sys embed vendor-controlled subdomains in the CN and SAN fields, offering a verifiable namespace for device identity.

By aligning with global PKI semantics, \sys enables scalable, standards-compliant device-to-device authentication, eliminating the need for vendor-specific authorization mechanisms or out-of-band pairing.



\begin{figure}[tbp]
    \centering
    \includegraphics[width=.95\linewidth]{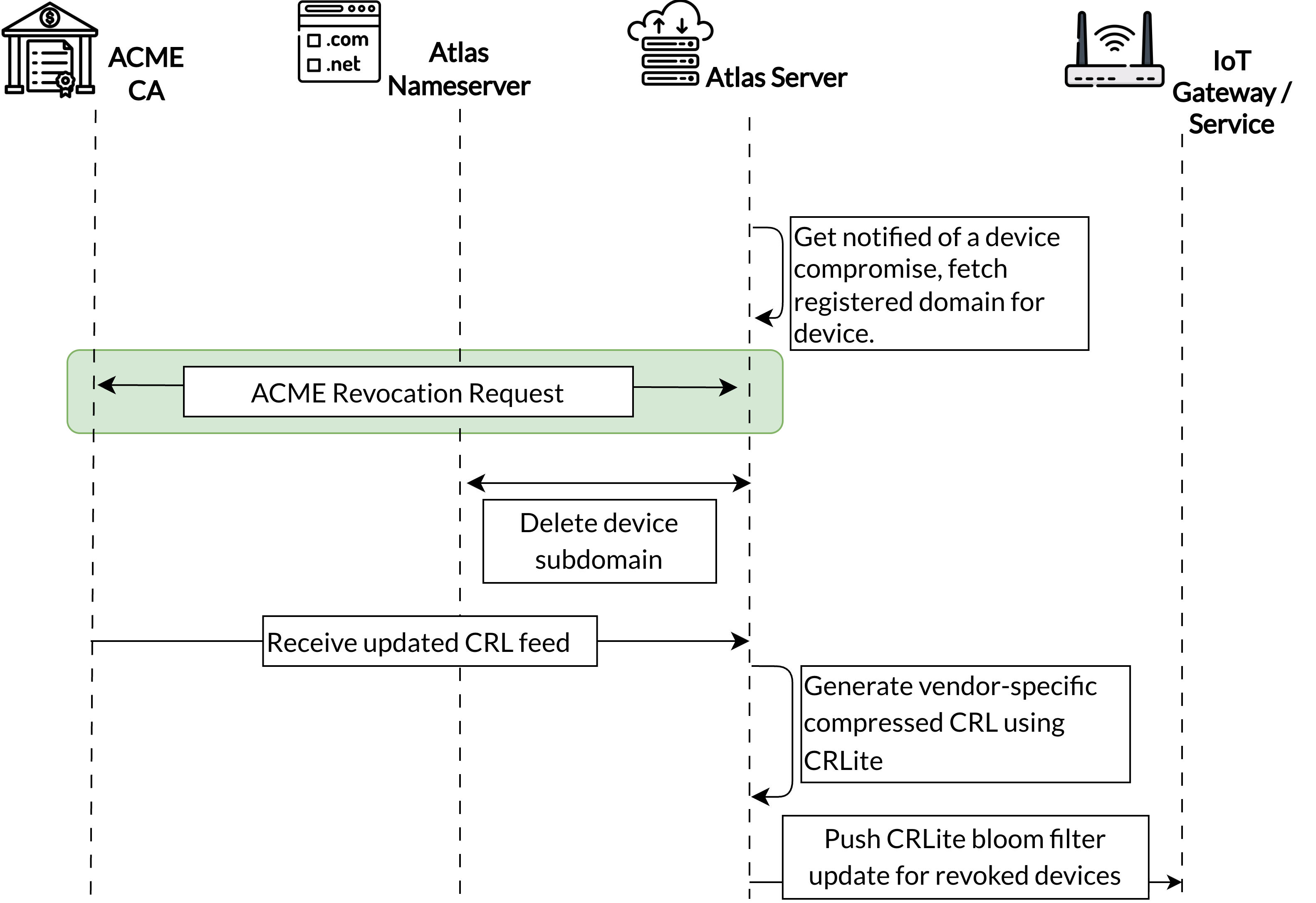}
    \caption{\sys revocation workflow: vendors periodically publish compressed revocation filters that encode revoked device certificates. Any IoT gateway or service that terminates device connections can subscribe to these filters and locally reject handshakes from compromised devices, without ever accessing device private keys.}
    \label{fig:revocation}
    \vspace{-1.5em}
\end{figure}

\section{Experimental Setup}
\label{sec:experimental-setup}

We implement \sys across three distinct components---IoT Device, IoT Cloud, and IoT Ecosystem---to comprehensively assess its impact across the entire IoT stack.

\paratitle{IoT Devices}
We use the Raspberry Pi 4 (RPi4)\cite{rpi4_specs}, Raspberry Pi Zero W (RPi0W)\cite{rpizerow_specs}, and ESP-WROOM-32 (ESP32)\cite{esp32_specs} development boards for our device experiments. These devices represent varying degrees of resource constraints, from multi-core processors to ultra-low-power microcontrollers. Importantly, all three devices support IEEE 802.11 wireless networking, which remains the dominant connectivity standard across consumer and industrial IoT deployments. These devices demonstrate a representative spectrum of real-world IoT hardware. Table~\ref{tab:device-characteristics} shows a comparison of the device capabilities.

\begin{table}[tb]
\centering
\small
\setlength{\tabcolsep}{2pt}
\caption{Device Characteristics}
\begin{tabular}{@{}l|l|l|l@{}}
\hline
\textbf{Characteristic} & \textbf{RPi4} & \textbf{RPi0W} & \textbf{ESP32} \\
\hline
Clock Speed & 1.4 GHz & 1.0 GHz & 240 MHz \\
RAM & 1 GB & 512 MB & 448 KB \\
OS Architecture & ARMv8 (64-bit) & ARMv6Z (32-bit) & Xtensa LX6 (32-bit) \\
WiFi Modes & 802.11ac & 802.11n  & 802.11b/g/n \\
\hline
\end{tabular}
\label{tab:device-characteristics}
\vspace{-1em}
\end{table}

\paratitle{IoT Cloud}
We deploy a custom IoT cloud instance on an Ubuntu 20.04 server that mimics a typical vendor-managed cloud. This deployment consists of standard web components, including an NGINX reverse proxy, a Flask-based web server, and a uWSGI application backend (Figure~\ref{fig:system-design}). To support \sys’s domain-bound identity model, we additionally configure an authoritative nameserver (using PowerDNS) responsible for managing the zone file associated with the vendor’s root domain. The zone file maintains DNS resource records—including A, CNAME, and TXT entries—that map device UUIDs to device-specific subdomains, as described in \xref{sec:domaing-binding}.

In contrast to our setup, most commercial IoT cloud deployments rely on recursive DNS resolvers (\eg provided by AWS Route 53 or Cloudflare DNS) and do not maintain their own authoritative DNS infrastructure. This limits the ability to programmatically manage per-device subdomain records or tightly control DNS-based identity bindings. By operating an authoritative DNS server within our IoT cloud instance, \sys enables dynamic and granular management of device-specific DNS entries.

For baseline comparisons, we also configure a deployment using AWS IoT Core, provisioning several Raspberry Pi 4 devices using the official AWS SDK\cite{aws2023iotcertificates}. AWS IoT provides out-of-the-box support for device provisioning, telemetry, and PKI services, serving as a representative model of cloud-mediated IoT infrastructure. This baseline allows us to contrast the operational and performance characteristics of traditional cloud-based device management with the \sys architecture.

\begin{figure}[tb]
   \centering
    \includegraphics[width=0.7\linewidth]{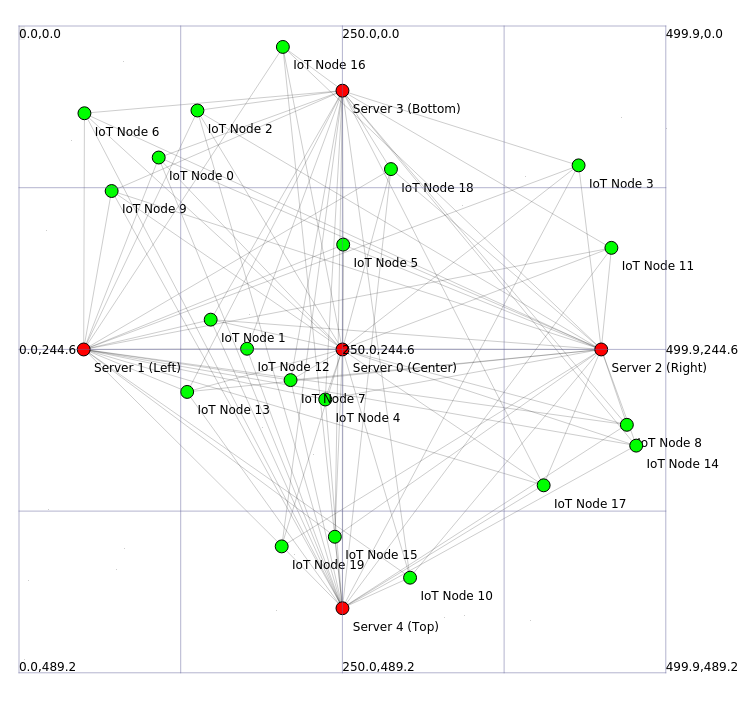}
    \caption{Smart City Network Simulation: Mobile IoT Nodes (\eg autonomous vehicles) interacting with static Servers (\eg traffic light systems) for mobility management or emergency response.}
    \label{fig:smart-city}
    \vspace{-2em}
\end{figure}

\begin{figure}[tb]
    \centering
    \includegraphics[width=0.8\linewidth]{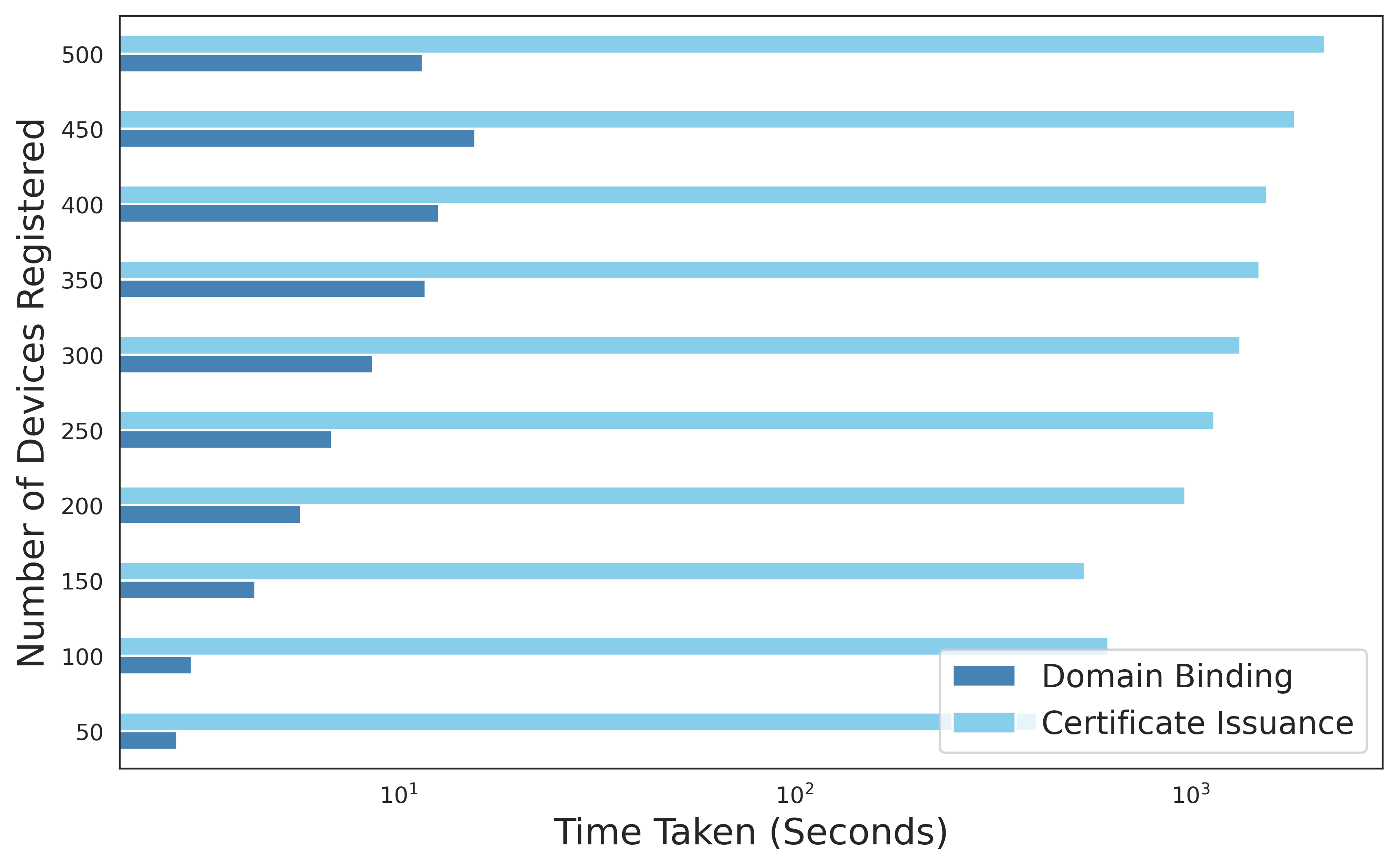}
    \caption{Bulk registration of devices using \sys is highly scalable.}
    \label{fig:bulk-registration}
\end{figure}

\begin{figure}[tb]
    \centering
    \includegraphics[width=0.8\linewidth]{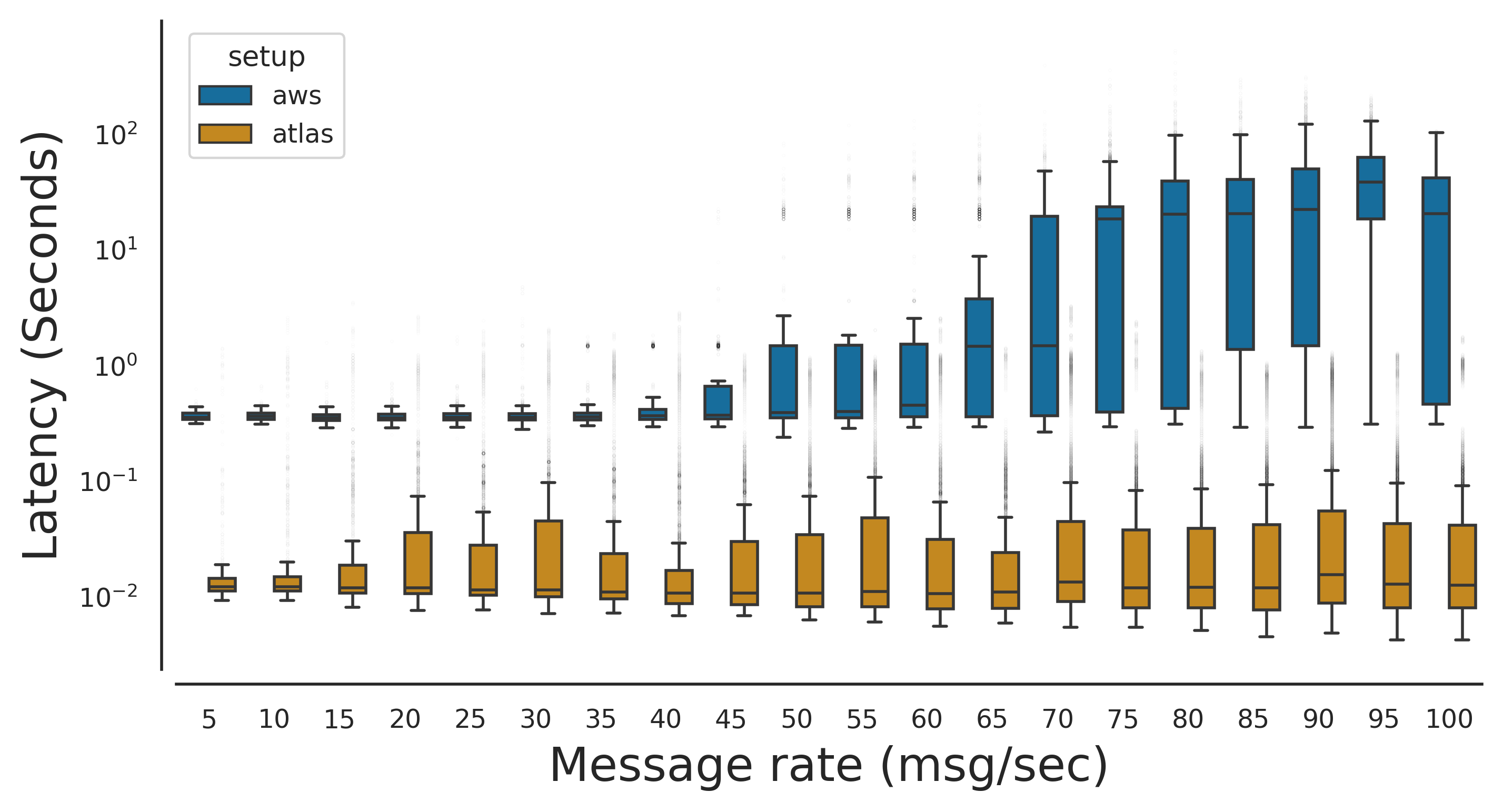}
    \caption{Real-world smart home applications developed using the \sys framework are consistently performant under scale workload and secure than the current setups that require mediation through IoT Clouds (AWS-IoT), which lead to high-variability in application performance.}
    \label{fig:cloud-vs-local-real-world}
    \vspace{-1em}
\end{figure}

\paratitle{IoT Ecosystem}
\emph{Smart Home Testbed ---} To demonstrate \sys in realistic deployment scenarios, we develop two testbeds. First, a smart home testbed consisting of heterogeneous IoT devices (RPi4, RPi0W, ESP32) connected through a commodity home router and coordinated via a local automation hub (using an MQTT broker \cite{mosquitto2023} on an RPi4). We contrast this with an AWS IoT-mediated setup, where two RPi4 devices, provisioned using AWS IoT Core, communicate through an IoT cloud (AWS IoT). We implemented cloud IoT applications using an AWS Lambda service that processes messages on the MQTT channels using AWS IoT Core. 

\emph{Smart City Testbed ---} One of the biggest challenges when designing IoT solutions is validating their applicability over large-scale deployments. It is impractical to build large scale testbeds with actual devices. Additionally, most commercial IoT devices are closed source, thus, modifying them for specific use cases is not viable. In order to validate the applicability of the \sys framework over large-scale deployments, such as smart cities, we leverage simulators such as ns-3~\cite{ns3_website}. ns-3 is one of the most widely used discrete-event network simulator for Internet systems, targeted primarily for research and educational purposes. We program a smart city model that reflects a realistic scenario where mobile IoT nodes (\eg autonomous vehicles) have to periodically interact with static IoT gateways (\eg traffic lights) spread throughout the city. Figure \ref{fig:smart-city} demonstrates a small scale simulation, for better visualization, where $15$ mobile IoT nodes are traversing through a $500mX500m$ city grid using the \emph{RandomWaypointMobilityModel} with speed varying from $5-25m/sec$. There are 5 statically allocated IoT gateways placed within the city which act as the aggregation nodes for device communications. Every mobile IoT node is programmed with a simple P2P application that looks for the nearest IoT gateway (within $100m$) every $2secs$, and starts sending a burst of messages to the gateway.

Under the \sys framework, these messages are transmitted to the gateway directly using secure D2D communication. However, under the current status quo of IoT cloud setups, this connection would be transmitted via vendor-specific clouds. We model this behavior using a simulated cloud latency derived using the Weibull distribution function defined in \ref{fig:weibull-formula}.